# Design and Analysis of Successive Decoding with Finite Levels for the Markov Channel

Teng Li, *Member, IEEE,* and Oliver M. Collins, *Fellow, IEEE*

*Abstract*— This paper proposes a practical successive decoding scheme with finite levels for the finite-state Markov channels where there is no a priori state information at the transmitter or the receiver. The design employs either a random interleaver or a deterministic interleaver with an irregular pattern and an optional iterative estimation and decoding procedure within each level. The interleaver design criteria may be the achievable rate or the extrinsic information transfer (EXIT) chart, depending on the receiver type. For random interleavers, the optimization problem is solved efficiently using a pilot-utility function, while for deterministic interleavers, a good construction is given using empirical rules. Simulation results demonstrate that the new successive decoding scheme combined with irregular low-density parity-check codes can approach the identically and uniformly distributed (i.u.d.) input capacity on the Markov-fading channel using only a few levels.

## I. INTRODUCTION

Many realistic communication systems suffer from unknown and time-varying channel conditions. The traditional strategy is single-code transmission and joint estimation and decoding. For example, iterative channel estimation and decoding was used in [1-3] for flat-fading channels and iterative equalization and decoding, so called turbo equalization, was used in [4] for inter-symbol interference (ISI) channels. With recent advances in low-density parity-check (LDPC) codes, channel estimation and decoding is combined into the message passing over a joint factor graph of the channel and the code, see [5] and [6] for block fading channels, [7] and [8] for ISI channels and [9] for Markov channels. In addition, codes need to be specifically optimized for the structure of the channel and the estimator. Various density evolution techniques have been proposed to find the optimal degree sequence of irregular LDPC codes for different channels [5-9]. Although shown to perform well for relatively short ISI channels [7], [8], this approach still has a performance gap for fading channels and Markov channels and its optimality for a general channel with memory is yet to be established.

This work was supported by the National Science Foundation under grants ECCS0523324, ECCS0801128 and CCF0830054. The material in this paper was presented in part at the IEEE International Symposium on Information Theory, Seattle, Washington USA, July 9 - July 14, 2006, and at the IEEE Information Theory Workshop, Lake Tahoe, California, September 2-6, 2007

Teng Li was with the Department of Electrical Engineering, University of Notre Dame, Notre Dame, IN 46556. He is now with Augusta Technology USA, Inc, Santa Clara, CA 95054. (email:teng.li@ieee.org).

Oliver M. Collins is with the Department of Electrical Engineering, University of Notre Dame, Notre Dame, IN 46556.

An alternative strategy is successive (or multistage) decoding with *multiple* codes. This technique uses a rectangular interleaver to multiplex $K$ independent codes into a single transmission stream at the transmitter and decodes them sequentially at the receiver. Successive decoding was originally developed to approach capacity for multilevel modulations [10], [11] and multiuser channels [12]. When applied to channels with memory, it effectively decomposes the physical channel into a bank of $K$ subchannels (levels) with weaker memory and additional decision feedback, where simplified algorithms, such as separate estimation and decoding (SED), and suboptimal codes, may perform well.

There has been extensive research on this topic. Pfister *et al.* first studied the achievable rate of successive decoding with SED for the ISI channel in [13] and the actual codes were then constructed by Soriaga *et al.* in [14]. Varnica *et al.* in [15] and Kavičić *et al.* in [16] adopted the successive decoding schedule for designing the component LDPC codes while performing the actual iterative decoding on the joint graph. A simplified scheme using only one estimator and one code of fixed rate for all subchannels was proposed by Narayanan and Nangare for ISI channels in [17] and by Li and Collins for correlated fading channels and other channels with memory in [18]. A pair of tight upper and lower bounds for the binary-input fading-channel capacity was derived in [19] and codes were designed to perform very close to the upper bound in [18] and [20].

Previous research has focused on asymptotic designs of the successive decoding. As the number of levels $K \to \infty$, the subchannels become memoryless and identical [17], [18] and the simple SED algorithm and the memoryless-channel optimized component code may have near-optimal performance on ISI channels [14], [17] and fading channels [18]. However, if a large number of codes (levels) are not allowed for practical reasons, the existing multi-rate designs in [13] and [14] and the rectangular interleavers in [14], [17], and [18] are no longer optimal.

This paper addresses the analysis and design of a more practical successive decoding scheme under the *finite*-level constraint. Since for a small $K$, the subchannels are no longer memoryless, we employ iterative estimation and decoding (IED) at each level to exploit the residual memory. More importantly, we propose an irregular interleaving pattern so that the $K$ codes may have different lengths and irregular symbol placements in the multiplexed transmission stream. In this configuration, the single-code iterative schemes [7-9] and those with pilot symbols [1-3,5,6] become two special cases of successive decoding of one and two levels, respectively. This framework offers more design freedom to tradeoff between performance and complexity. The deterministic irregular interleaver is difficult to optimize because there are $K^N$ possible patterns for an $N$-bit long $K$-level interleaver. Therefore, we propose a random interleaver specified by the weight distribution $\mathbf{w} = [w_1, \cdots, w_K]$, where $w_k \geq 0$ and $\sum_{k=1}^{K} w_k = 1$. It acts as a multiplexer that chooses a bit from code $k$ to transmit with probability $w_k$. The resulting $N$-bit sequence is expected to have approximately $w_k N$ bits from level $k$ at random positions. The random interleaver is much easier to design and is asymptotically optimal. It also provides a new interpretation of the area property of the extrinsic information transfer (EXIT) function [21].

This paper develops the successive decoding scheme for finite-state Markov channels (FSMCs) whose state evolves independently of the channel input. FSMCs are good approximations to many realistic channels and have been extensively investigated [22]. For simplicity we consider the identically and uniformly distributed (i.u.d.) binary channel input. Consequently, we use the maximal achievable information rate when the channel input is an i.u.d. binary sequence as performance measure. This information rate is called the i.u.d. capacity $C^{\text{i.u.d.}}$ in this paper following the notations in [7] and [8]. It is also known as the symmetric information rate [13]. In order to achieve the channel capacity $C \geq C^{\text{i.u.d.}}$, the i.u.d. binary sequence may be passed through a nonlinear device with memory, such as the inner nonlinear trellis encoder [16], to mimic the optimal distribution. The designs in this paper are readily applicable to the concatenation of this nonlinear device and the original physical channel as well as other channels with memory after modifying the estimator.

The proposed successive decoding technique is analyzed by comparing the achievable rate (of the SED algorithm) to the i.u.d. capacity. After expressing the rate difference in terms of the state-transition matrix of the underlying Markov channel, we show that the achievable rate goes to the i.u.d. capacity exponentially fast as $K \to \infty$ for both the rectangular interleaver and the equal-weighted random interleaver. However, when $K$ is small, these two rates diverge. The difference between them comes from the mutual information loss caused by the memoryless-channel assumption in the SED and can be recovered with a more sophisticated receiver. In the literature, two conceptually different Monte-Carlo methods are used to estimate the i.u.d. capacity [13], [23] and the achievable rate [13], [8]. We propose a unified way to estimate both of them from the output *a posteriori* probability of the BCJR algorithm [24] by changing the distribution of the *a priori* information.

The mutual information analysis is used for system design as well. The design objective is to maximize the supported information rate of a finite-level successive decoding scheme through interleaver optimization. We will adopt the existing set of AWGN-channel optimized LDPC codes [25] and make no attempt to optimize them for the specific channel and receiver. The code rate, however, needs to be properly chosen according to some mutual information measure at each level. For the SED, we use the achievable rate, while for the IED, we use the maximal code rate at which the EXIT chart analysis [26] still predicts the convergence of the iterative process. These rates also become the objective functions in optimization.

For random interleavers, both the achievable rate and the EXIT function can be efficiently estimated from a so-called pilot-utility function, which measures the achievable rate of a single-code system as a function of the percentage of randomly-placed pilot symbols. We show that the subchannel achievable rate is a point on the pilot-utility function and the overall achievable rate is the area of a $K$-step piecewise constant curve beneath it. The EXIT function of the subchannel estimator is simply the transformation of a segment on the pilot-utility function. Consequently, the achievable rate of the SED algorithm can be maximized semi-analytically by a recursive method. The optimal weight distribution of the random interleaver under the IED is also found by matching a set of subchannel EXIT functions to the set of decoder EXIT functions so that the overall code rate is maximized. Essentially, instead of designing codes for the channel according to the traditional wisdom [5-9], we design the subchannel to match the code through interleaver optimization. However, for deterministic interleavers, computationally intensive Monte-Carlo simulation is required to estimate the mutual information. Hence, we will only optimize a class of interleavers that are empirically good.

If a more stringent overall-delay constraint is imposed, the effect of the finite codeword length must be considered. As the number of levels increases, the achievable rate will increase but the code lengths and their performance will decrease. There is an optimal number of levels for a given delay constraint. We use the random-coding bound [27] that relates the word-error probability to the codeword length to study this trade-off. Numerical results show that for the example channel, it is beneficial to use more than two levels, provided that a moderate delay of several thousand symbols is allowed.

The outline of the paper is as follows. Section II introduces the basic concepts of successive decoding. In Section III, we present the definitions of i.u.d. capacity and achievable rate of the subchannel and show the convergence of the achievable rate to the i.u.d. capacity for the rectangular interleaver. Section IV deals with the design of interleavers for the SED. We first discuss the properties and the asymptotic optimality of random interleavers. We then introduce the pilot-utility function and its properties and apply them for interleaver optimization. A set of good deterministic interleavers are also given. Section V proposes the EXIT chart analysis of the IED algorithm for system design. Section VI uses the random-coding error exponents to analyze the impact of a finite delay constraint. Section VII presents some numerical results and Section VIII concludes the paper.

## II. Successive Decoding

### A. Channel Model

This paper considers a Markov-modeled flat-fading channel with additive white Gaussian noise (AWGN). The received signal $Y_t$ is given by

$$Y_t = H_t X_t + W_t \tag{1}$$

where $H_t$, $X_t$, and $W_t$ are the complex channel gain unknown to both the receiver and the transmitter, the transmitted symbol,

and the AWGN, respectively, and are assumed to be mutually independent. The input is an independently and uniformly distributed (i.u.d.) binary sequence, $X_t \in \{-1, +1\}$, with power $E_s = 1$. The AWGN has a symmetric complex Gaussian distribution, $W_t \sim \mathcal{CN}(0, N_0)$. The channel state process forms an irreducible, aperiodic, and stationary Markov chain over a finite state space $H_t \in \{A_1, \cdots, A_Q\}$ with state transition probability

$$P_{q',q} = \Pr(H_t = A_{q'} | H_{t-1} = A_q) \qquad (2)$$

and stationary state probability

$$P_q = \Pr(H_t = A_q)$$

where $q, q' = 1, \cdots, Q$. We call $P = [P_{q',q}]_{q',q=1,\cdots,Q}$ the state-transition matrix. We use upper-case letters for random variables, lower-case ones for their realizations, and boldface letters for vectors. We use the notation $\Pr()$ for both the probability mass function and the probability density function (PDF).

### B. Encoding

In a $K$-level successive decoding scheme, the transmitter partitions the information bits into $K$ sub-sequences according to an interleaving pattern, and independently encodes them into $K$ codewords of length $N_k$ bits and rate $r_k$ for $k = 1, \cdots, K$. Let $\mathbf{x}_k = \{x_{1,k}, \cdots, x_{N_k,k}\}$ denote the codeword $k$. A total of $N = \sum_{k=1}^{K} N_k$ bits from $\mathbf{x}_1, \cdots, \mathbf{x}_K$ are then interleaved into a transmission stream $\{x_t\}_{t=1}^{N}$ with an overall rate of $r = \sum r_k N_k / N$. We assume the limit $w_k = \lim_{N \to \infty} N_k / N$, called the *weight* of the $k$th level, exists.

We assume there is another interleaving (deinterleaving) mechanism embedded at the encoder (decoder). Hence the interleaver considered here does not scramble the bits within an individual codeword and may be represented by a vector

$$\boldsymbol{\pi} = [\pi_1, \cdots, \pi_N], \quad \pi_t \in \{1, \cdots, K\} \text{ for } t = 1, \cdots, N.$$

This means a bit from the codeword $\pi_t$ is transmitted at time instant $t$ so that $x_t = x_{i,\pi_t}$ for some $i$. The one-to-one mapping between the pair of indices $(i, k)$ and the time index $t$ at which $x_{i,k}$ is transmitted is conveniently represented by a function $t = t(i, k)$ so that $x_{i,k} = x_{t(i,k)}$. For example, if the interleaver is $\boldsymbol{\pi} = [1, 3, 2, 3, 2, 3]$, then the transmitted sequence $[x_1, \cdots, x_6] = [x_{1,1}, x_{1,3}, x_{1,2}, x_{2,3}, x_{2,2}, x_{3,3}]$, and $t(1,1) = 1$, $t(1,2) = 3$, and $t(2,2) = 5$ and so on.

The interleaver configuration is key to the successive decoding design since it determines the channel configuration for each codeword. Both a deterministic and a random interleaver construction will be considered here.

*Definition 1:* A $K$-level *deterministic* interleaver is constructed from the repetition of a subpattern $\boldsymbol{\omega} \in \mathbb{K}^L$ of a fixed length $L$ so that $\boldsymbol{\pi} = [\boldsymbol{\omega}, \cdots, \boldsymbol{\omega}] \in \mathbb{K}^N$, where $\mathbb{K} = \{1, \cdots, K\}$.

A simple deterministic interleaver is the rectangular interleaver with $\boldsymbol{\omega} = [1, \cdots, K]$. It is asymptotically optimal [18] as $K \to \infty$ but the free parameter $K$ may be very large for good performance. A more general deterministic interleaver, called the *irregular* interleaver [28], is a permutation $\boldsymbol{\omega} = \text{perm}(\mathbf{1}_{L_1}, \mathbf{2}_{L_2}, \cdots, \mathbf{K}_{L_K}) \in \mathbb{K}^L$, where $\mathbf{k}_n = [k, \cdots, k]$ is a row vector of length $n$ and $\sum_{k=1}^{K} L_k = L$. Here perm denotes a permutation function. The pilot-symbol assisted modulation (PSAM) can be viewed as a special case of successive decoding with only two levels i.e. $\boldsymbol{\omega} = [1, 2, \cdots, 2]$.

Contrary to the rectangular interleaver, the irregular interleaver has a design space of $K^L$ possible subpatterns $\boldsymbol{\omega}$, making optimization difficult. This partially motivates a class of random interleaver defined as follows.

*Definition 2:* A $K$-level *random* interleaver of weight distribution $\mathbf{w} = [w_1, \cdots, w_K]$ is a random vector $\boldsymbol{\Pi} = [\Pi_1, \cdots, \Pi_N]$ whose entries are identically-and-independently distributed (i.i.d.) according to the probability mass function $\Pr(\Pi_t = k) = w_k$ for $1 \leq t \leq N$ and $1 \leq k \leq K$, where $w_k \geq 0$ and $\sum_{k=1}^{K} w_k = 1$.

The random interleaver is completely specified by the weight distribution. Its optimization can be carried out over a $K$ dimensional space and is much more feasible. Its properties will be discussed in Section IV.

### C. Estimation and Decoding

In successive decoding, the $K$ codewords are decoded one by one. Their hard-decisions are fed back to subsequent levels and are treated as known training symbols. Thus more training symbols are produced as decoding proceeds to higher levels. Each stage performs either *separate* estimation and decoding (SED) or *iterative* estimation and decoding (IED). The first approach estimates the codeword symbols over the trellis of the underlying FSMC, and then invokes the decoder with the set of likelihood ratios as its sole input. By separating the two processes, it ignores the interaction between the channel memory and the code structure and thus has a performance penalty. Nonetheless, SED for a deep rectangular interleaver is shown in [14], [17], and [18] to be i.u.d. capacity approaching because the underlying subchannel tends to be memoryless as the number of levels goes to infinity. However, when the design has finite levels, the codeword symbols may be closely placed. It is then necessary to address jointly the memory of both the channel and the encoder. The IED is a computationally efficient way to do so and is widely used in the literature, see for example [1-4].

Consider the decoding process of the codeword $\mathbf{x}_k$. Assume codewords $\mathbf{x}_1$ to $\mathbf{x}_{k-1}$ have been correctly decoded and become the set of known training symbols denoted by a sequence $\mathbf{u}_k = [u_1, \cdots, u_N]$ where

$$u_t = \begin{cases} x_t, & \text{if } x_t \text{ is from codeword } \mathbf{x}_1 \text{ to } \mathbf{x}_{k-1} \\ \phi, & \text{otherwise.} \end{cases}$$

Here $\phi$ denotes an erased symbol. The receiver estimates the channel state along the trellis of the FSMC using a BCJR algorithm similar to [24]. For notational convenience, we introduce a windowing operation $\langle \cdot \rangle_{t_1}^{t_2}$, where $t_1$ and $t_2$ are the start and end time of the window, respectively. Suppose $\mathbf{a} = [a_1, \cdots, a_N]$ is a vector indexed by time, then $\langle \mathbf{a} \rangle_{t_1}^{t_2} = [a_{t_1}, \cdots, a_{t_2}]$. Similarly, the windowing operation on the codeword $\mathbf{X}_k = [X_{1,k}, \cdots, X_{N_k,k}]$ yields $\langle \mathbf{X}_k \rangle_{t_1}^{t_2} = [X_{i_1,k}, \cdots, X_{i_2,k}]$, where $t(i_1 - 1, k) < t_1 \leq t(i_1, k)$ and

$t(i_2, k) \leq t_2 < t(i_2 + 1, k)$. Let the forward and backward state probabilities, respectively, be

$$\alpha_t(q) = \Pr(H_t = A_q, \langle \mathbf{y} \rangle_1^{t-1}, \langle \mathbf{u}_k \rangle_1^{t-1}) \quad (3)$$
$$\beta_t(q) = \Pr(\langle \mathbf{y} \rangle_t^N, \langle \mathbf{u}_k \rangle_t^N | H_t = A_q). \quad (4)$$

Define the branch metric to reflect whether a known symbol is present at time instant $t$ as

$$\gamma_t(q', q)$$
$$= \begin{cases} \Pr(y_t, x_t, H_{t+1} = A_{q'} | H_t = A_q), & \text{if } x_t \text{ is known} \\ \Pr(y_t, H_{t+1} = A_{q'} | H_t = A_q), & \text{otherwise} \end{cases}$$
$$= \begin{cases} \gamma_t(q', q, x_t), & \text{if } x_t \text{ is known} \\ \sum_{X_t = \pm 1} \gamma_t(q', q, X_t), & \text{otherwise} \end{cases} \quad (5)$$

where $\gamma_t(q', q, X_t)$ is the conditional branch metric given by

$$\gamma_t(q', q, X_t)$$
$$= \Pr(y_t, X_t, H_{t+1} = A_{q'} | H_t = A_q) \quad (6)$$
$$= \Pr(X_t) \Pr(H_{t+1} = A_{q'} | H_t = A_q) \Pr(y_t | H_t = A_q, X_t) \quad (7)$$
$$= \Pr(X_t) P_{q', q} \exp\left(-|y_t - X_t A_q|^2 N_0^{-1}\right) (\pi N_0)^{-1}. \quad (8)$$

In the above, $\Pr(X_t) = 1/2$ for $X_t = \pm 1$ is the *a priori* probability and (7) is from $\Pr(y_t, X_t | H_{t+1}, H_t) = \Pr(X_t) \Pr(y_t | X_t, H_t)$ since $H_t$ evolves independently of $X_t$ and $y_t$ only conditionally depends on $H_t$ and $X_t$. The $\alpha_t(q)$ and $\beta_t(q)$ are computed recursively as

$$\alpha_t(q) = \sum_{q'=1}^Q \gamma_{t-1}(q, q') \alpha_{t-1}(q'), \quad t = 2, \cdots, N \quad (9)$$
$$\beta_t(q) = \sum_{q'=1}^Q \gamma_t(q', q) \beta_{t+1}(q'), \quad t = N, \cdots, 1 \quad (10)$$

where $\alpha_1(q) = P_q$ and $\beta_{N+1}(q) = P_q$ for $1 \leq q \leq Q$. The estimator computes the likelihood ratio of each bit in the codeword $k$ as

$$\Lambda^e(X_t = a) = \frac{\Pr(X_t = a, \mathbf{y}, \mathbf{u}_k)}{\Pr(X_t = -a, \mathbf{y}, \mathbf{u}_k)}$$
$$= \frac{\sum_{q=1}^Q \sum_{q'=1}^Q \alpha_t(q) \gamma_t(q', q, a) \beta_{t+1}(q')}{\sum_{q=1}^Q \sum_{q'=1}^Q \alpha_t(q) \gamma_t(q', q, -a) \beta_{t+1}(q')} \quad (11)$$

for $t = t(1, k), \cdots, t(N_k, k)$. The decoder treats the sequence of likelihood ratios $\{\Lambda^e(X_{i,k} = +1)\}_{i=1}^{N_k}$ as i.i.d. samples from a memoryless channel and decodes $\mathbf{x}_k$ accordingly.

The IED scheme exchanges the extrinsic soft information repetitively between the estimator and the decoder within each individual subchannel. Consider the $n$th iteration at the $k$th subchannel. Suppose the decoder output likelihood ratio at the $(n-1)$th iteration is $\Lambda_{n-1}^d(X_{i,k})$, its extrinsic output is

$$L_{n-1}^{d,out}(X_{i,k}) = \Lambda_{n-1}^d(X_{i,k}) / L_{n-1}^{d,in}(X_{i,k}) \quad (12)$$

for $i = 1, \cdots, N_k$, where $L_{n-1}^{d,in}(X_{i,k})$ is the decoder extrinsic input. The estimator then computes its output likelihood ratio $\Lambda_n^e(X_{i,k})$ according to (9) - (11), except that the *a priori* probability $\Pr(X_t)$ in (8) is replaced by its extrinsic input $L_n^{e,in}(X_{i,k}) = L_{n-1}^{d,out}(X_{i,k})$, where $t = t(i, k)$. The estimator output extrinsic information at the $n$th iteration is

$$L_n^{e,out}(X_{i,k}) = \Lambda_n^e(X_{i,k}) / L_n^{e,in}(X_{i,k}) \quad (13)$$

for $i = 1, \cdots, N_k$, which becomes the decoder extrinsic input $L_n^{d,in}(X_{i,k}) = L_n^{e,out}(X_{i,k})$. The above process starts with the initial condition $L_0^{e,in}(X_{i,k}) = 1$ and repeats until a stopping criterion is satisfied. For example, the parity-check equations hold for an LDPC decoder or the maximum number of iterations is reached.

## III. MUTUAL INFORMATION

This section introduces two mutual information: the achievable rate $R$ and the i.u.d. capacity $C^{\text{i.u.d.}}$. Both of them are derived for the i.u.d. binary channel inputs. While $C^{\text{i.u.d}}$ denotes the maximal achievable information rate given any receiver, $R$ denotes the achievable information rate of the SED algorithm. Note the channel capacity $C \geq C^{\text{i.u.d}}$ requires an optimal input distribution. The distance between $R$ and $C^{\text{i.u.d}}$ is then used to show how fast a finite-level rectangular interleaver based scheme converges.

### A. The Achievable Rate and the i.u.d. Capacity

Analogous to a multiuser system [29], the $K$ independent codewords in a successive decoding scheme with perfect decision feedback[1] are equivalently transmitted over $K$ parallel subchannels (also called equivalent channels [11]). The $k$th subchannel is defined as a channel with a vector input $\mathbf{X}_k$, a vector output $\mathbf{Y}$, a training sequence $\mathbf{U}_k$, and a channel transition probability $\Pr(\mathbf{Y}|\mathbf{X}_k, \mathbf{U}_k)$. For either the deterministic or the random interleavers, the subchannel is a stationary, ergodic, and indecomposable [27] FSMC. Following [27], we define the i.u.d. capacity of each subchannel as follows.

*Definition 3:* The i.u.d. capacity of the $k$th subchannel is the mutual information between the i.u.d. input vector $\mathbf{X}_k$ and the output vector $\mathbf{Y}$ conditioned on the training sequence $\mathbf{U}_k$

$$C_k^{\text{i.u.d.}} = \lim_{N \to \infty} \frac{1}{N_k} I(\mathbf{X}_k; \mathbf{Y} | \mathbf{U}_k) \quad (14)$$

and the i.u.d. capacity of the physical channel (1) is the mutual information between the i.u.d. input vector $\mathbf{X}$ and the output vector $\mathbf{Y}$

$$C^{\text{i.u.d.}} = \lim_{N \to \infty} \frac{1}{N} I(\mathbf{X}; \mathbf{Y}). \quad (15)$$

Applying the chain rule of mutual information to (15) yields

$$C^{\text{i.u.d.}} = \sum_{k=1}^K w_k C_k^{\text{i.u.d.}}. \quad (16)$$

Achieving (14) with one code would require a joint maximum-likelihood decoder. On the other hand, the simple SED over a subchannel can achieve the following rate.

---
[1]Throughout the paper, the decision feedback is assumed to be perfect. The effects of imperfect decisions are minimal in the proposed successive decoding because decisions are generated by a strong component code, which, when operating at the designed region, will produce a small BER that has little effect on the estimation process in the later stage. A more detailed discussion on the imperfect decision feedback can be found in [18].

*Definition 4:* The achievable rate of the $k$th subchannel with SED is the average of the conditional mutual information between the individual bit $X_{i,k}$ and the channel output $\mathbf{Y}$

$$R_k = \lim_{N \to \infty} \frac{1}{N_k} \sum_{i=1}^{N_k} I(X_{i,k}; \mathbf{Y}|\mathbf{U}_k) \quad (17)$$

and the overall achievable rate of the physical channel (1) is

$$R = \sum_{k=1}^{K} w_k R_k. \quad (18)$$

The achievable rate is the maximal rate of error-free communication with the suboptimal decoding rule [30] that approximates the true APP $\Pr(\mathbf{X}_k|\mathbf{Y},\mathbf{U}_k)$ by a product $\prod \Pr(X_{i,k}|\mathbf{Y},\mathbf{U}_k)$ under the (memoryless subchannel) assumption that $\{X_{i,k}\}_{i=1}^{N_k}$ are conditionally independent. It is a lower bound of the subchannel i.u.d. capacity since

$$C_k^{\text{i.u.d.}} - R_k$$
$$= E_{\Pr(\mathbf{X}_k|\mathbf{Y},\mathbf{U}_k)} \left[ \log_2 \frac{\Pr(\mathbf{X}_k|\mathbf{Y},\mathbf{U}_k)}{\prod_{i=1}^{N_k} \Pr(X_{i,k}|\mathbf{Y},\mathbf{U}_k)} \right]$$
$$= D_{KL}\Big( \Pr(\mathbf{X}_k|\mathbf{Y},\mathbf{U}_k) \, \Big\| \, \prod_{i=1}^{N_k} \Pr(X_{i,k}|\mathbf{Y},\mathbf{U}_k) \Big)$$
$$\geq 0 \quad (19)$$

where (19) is due to the non-negativity of the Kullback-Leibler distance $D_{KL}$ [29]. The information rate (17) was also derived in [8] as the performance bound of decoding LDPC codes over ISI channels with a single pass of the BCJR algorithm.

*B. Estimation of the Achievable Rate and the i.u.d. Capacity*

From the definition of the mutual information and the entropy [29], expression (17) can be calculated as follows

$$R_k = \lim_{N \to \infty} \frac{1}{N_k} \sum_{i=1}^{N_k} \Big( H(X_{i,k}) - H(X_{i,k}|\mathbf{Y},\mathbf{U}_k) \Big)$$
$$= 1 - \lim_{N \to \infty} \frac{1}{N_k} \sum_{i=1}^{N_k} E\Big[$$
$$\quad - \log_2 \Pr(X_{i,k} = x_{i,k} | \mathbf{Y} = \mathbf{y}, \mathbf{U}_k = \mathbf{u}_k) \Big] \quad (20)$$
$$= 1 - \lim_{N \to \infty} \frac{1}{N_k} \sum_{i=1}^{N_k} E\left[ -\log_2 \frac{\Lambda^e(x_{i,k})}{1 + \Lambda^e(x_{i,k})} \right]. \quad (21)$$

The expectation in (21) can be evaluated through a Monte-Carlo integration, which simulates the channel output and computes $\{\Lambda^e(x_{i,k})\}$ using the BCJR algorithm according to (9) - (11). In fact, the computation of $R_k$ directly corresponds to the estimation process in the SED, except that the likelihood ratio of the actual input realization $x_{i,k}$ must be used.

Applying the chain rule of mutual information to (14) yields

$$C_k^{\text{i.u.d.}} = \lim_{N \to \infty} \frac{1}{N_k} \sum_{i=1}^{N_k} I(X_{i,k}; \mathbf{Y}|\mathbf{U}_k, X_{1,k}, \cdots, X_{i-1,k}). \quad (22)$$

According to (22), we introduce an additional training sequence $\{x_{1,k}, \cdots, x_{i-1,k}\}$ in the forward recursion of the BCJR algorithm to compute a new likelihood ratio of $x_{i,k}$ as

$$\widetilde{\Lambda}^e(x_{i,k}) = \frac{\Pr(x_{i,k}, \mathbf{y}, \mathbf{u}_k, x_{1,k}, \cdots, x_{i-1,k})}{\Pr(-x_{i,k}, \mathbf{y}, \mathbf{u}_k, x_{1,k}, \cdots, x_{i-1,k})} \quad (23)$$

for $i = 1, \cdots, N_k$. Hence

$$C_k^{\text{i.u.d.}} = 1 - \lim_{N \to \infty} \frac{1}{N_k} \sum_{i=1}^{N_k} E\left[ -\log_2 \frac{\widetilde{\Lambda}^e(x_{i,k})}{1 + \widetilde{\Lambda}^e(x_{i,k})} \right]. \quad (24)$$

Note combining (16) and (24) yields the i.u.d. capacity of the physical channel with memory. This is an alternative to the methods presented in [23] and [13] and was hinted, though not pursued, in [23].

*C. Convergence of the Achievable Rate to the i.u.d. Capacity*

Under the mild conditions of positive state-transition matrix and noisy channel output, a hidden Markov model has exponential decay of the channel memory such that the difference between the state estimates at time $t$ with or without the initial channel knowledge at $t-n$ goes to zero exponentially fast with respect to $n$ due to state mixing, see for example [31] and [32]. This implies that the increase in mutual information between $X_t$ and $\mathbf{Y}$ due to some additional training symbols located at least $n$ symbols away also goes to zero at an exponential rate as shown in Lemma 1.

*Lemma 1:* Assume that the channel state-transition matrix $P$ defined in (2) is primitive, i.e., $P_{q',q} > 0$ and assume that $\Pr(y_t|H_t = A_q, X_t) > 0$ for any $1 \leq q \leq Q$, $1 \leq q' \leq Q$, $X_t \in \{-1, +1\}$, and $1 \leq t \leq N$. The conditional mutual information $I(X_t; \langle \mathbf{Y} \rangle_{t-m}^{t+n} | \langle \mathbf{U} \rangle_{t-m}^{t+n})$, where $\mathbf{U}$ is an arbitrary sequence of training symbols, will converge exponentially fast with respect to $m$ for any $n \geq 0$ so that

$$I\big(X_t; \langle \mathbf{Y} \rangle_{t-m'}^{t+n} | \langle \mathbf{U} \rangle_{t-m'}^{t+n}\big) - I\big(X_t; \langle \mathbf{Y} \rangle_{t-m}^{t+n} | \langle \mathbf{U} \rangle_{t-m}^{t+n}\big)$$
$$< \ln(2)^{-1} \max_{1 \leq i \leq Q, 1 \leq j \leq Q} d(\mathbf{p}_i, \mathbf{p}_j) \tau(P)^{m-1}$$

where $m' > m > 0$. Here $\mathbf{p}_i$ for $1 \leq i \leq Q$ is the column vectors of $P$, $0 \leq \tau(P) < 1$ is the Birkhoff contraction coefficient of a strictly positive matrix $P$ defined as

$$\tau(P) = \sup_{\mathbf{u}>0, \mathbf{v}>0, \mathbf{u} \neq \lambda \mathbf{v}} \frac{d(P\mathbf{u}, P\mathbf{v})}{d(\mathbf{u}, \mathbf{v})} \quad (25)$$

and

$$d(\mathbf{u}, \mathbf{v}) = \ln \max_{i,j} \left( \frac{u(i)}{v(i)} \frac{v(j)}{u(j)} \right) \geq 0 \quad (26)$$

is the Hilbert metric between two positive vectors $\mathbf{u}$ and $\mathbf{v}$.

*Proof:* See appendix I. ∎

The statement and the proof of Lemma 1 involve the theory of the product of positive matrices, see [31], [33], and [34]. The properties of both the Hilbert metric and the Birkhoff contraction coefficient can be found in [31].

Applying Lemma 1 to the rectangular interleaver, the following theorem shows that, as $K \to \infty$, $\Pr(\mathbf{X}_k|\mathbf{Y},\mathbf{U}_k) \to \prod_{i=1}^{N_k} \Pr(X_{i,k}|\mathbf{Y},\mathbf{U}_k)$ and the subchannel converges to a memoryless channel where the SED is indeed optimal.

*Theorem 1:* Under the same assumptions of Lemma 1, the achievable rate $R$ of a $K$-level rectangular interleaver approaches $C^{\text{i.u.d.}}$ exponentially fast with respect to $K$, so

$$C^{\text{i.u.d.}} - R < \ln(2)^{-1} \max_{1 \leq i \leq Q, 1 \leq j \leq Q} d(\mathbf{p}_i, \mathbf{p}_j) \tau(P)^{K-2} \quad (27)$$

where $P$ is the state-transition matrix (2), $\mathbf{p}_i$ is the column vector of $P$, and $0 \leq \tau(P) < 1$ and $d(\mathbf{p}_i, \mathbf{p}_j) \geq 0$ are the Birkhoff contraction coefficient (25) and the Hilbert metric (26), respectively.

*Proof:* From (22) and (17), the difference between $C_k^{\text{i.u.d.}}$ and $R_k$ is upper bounded by

$$C_k^{\text{i.u.d.}} - R_k$$
$$= \lim_{N \to \infty} \frac{1}{N_k} \sum_{i=1}^{N_k} \Big( I(X_{i,k}; \mathbf{Y} | \mathbf{U}_k, X_{1,k}, \cdots, X_{i-1,k})$$
$$- I(X_{i,k}; \mathbf{Y} | \mathbf{U}_k) \Big) \quad (28)$$
$$\leq \lim_{N \to \infty} \frac{1}{N_k} \sum_{i=1}^{N_k} \Big( I(X_{i,k}; \mathbf{Y} | \mathbf{U}_k, X_{1,k}, \cdots, X_{i-1,k})$$
$$- I(X_{i,k}; \langle \mathbf{Y} \rangle_{t(i,k)-K+1}^N | \langle \mathbf{U}_k \rangle_{t(i,k)-K+1}^N) \Big) \quad (29)$$
$$\leq \ln(2)^{-1} \max_{1 \leq i \leq Q, 1 \leq j \leq Q} d(\mathbf{p}_i, \mathbf{p}_j) \tau(P)^{K-2} \quad (30)$$

for $k = 1, \cdots, K$, where (29) is true because eliminating some channel outputs and training symbols reduces mutual information and (30) is a result of Lemma 1. The result (27) follows from (30) and the fact that $C^{\text{i.u.d.}} = \sum_{k=1}^K C_k^{\text{i.u.d.}}/K$ and $R = \sum_{k=1}^K R_k/K$. ∎

## IV. DESIGN FOR SEPARATE ESTIMATION AND DECODING

In this section, we first show some properties of the random interleaver and then address the random interleaver optimization problem using a pilot-utility function. We also give a good deterministic interleaver constructed according to empirical rules.

### A. Properties of the Random Interleaver

Under a random interleaver, the configuration of a subchannel as well as its achievable rate and i.u.d. capacity depend on the realization of the interleaver. We define the (ensemble) average of $C_k^{\text{i.u.d.}}$ and the (ensemble) average of $R_k$, respectively, as

$$\mathbb{C}_k^{\text{i.u.d.}} = \lim_{N \to \infty} E_{\mathbf{\Pi}} \left[ \frac{1}{N_k} I_{\mathbf{\Pi}}(\mathbf{X}_k; \mathbf{Y} | \mathbf{U}_k) \right] \quad (31)$$

$$\mathbb{R}_k = \lim_{N \to \infty} E_{\mathbf{\Pi}} \left[ \frac{1}{N_k} \sum_{i=1}^{N_k} I_{\mathbf{\Pi}}(X_{i,k}; \mathbf{Y} | \mathbf{U}_k) \right] \quad (32)$$

where the subscript $\mathbf{\Pi}$ is introduced in the mutual information to denote its dependency on the underlying interleaver pattern. The (ensemble) average of $C^{\text{i.u.d.}}$ and the (ensemble) average of $R$ are respectively

$$\mathbb{C}^{\text{i.u.d.}} = \sum_{k=1}^K w_k \mathbb{C}_k^{\text{i.u.d.}} \quad (33)$$

and

$$\mathbb{R} = \sum_{k=1}^K w_k \mathbb{R}_k. \quad (34)$$

The following proposition shows that, as the length of a random interleaver realization goes to infinity, both the i.u.d. capacity and the achievable rate of a subchannel converge to their ensemble averages in probability.

*Proposition 1:* For a random interleaver $\mathbf{\Pi}$ with i.i.d. entries, as $N \to \infty$

$$I_{\mathbf{\Pi}}(\mathbf{X}_k; \mathbf{Y} | \mathbf{U}_k)/N_k \xrightarrow{\text{p}} \mathbb{C}_k^{\text{i.u.d.}} \quad (35)$$

$$\sum_{i=1}^{N_k} I_{\mathbf{\Pi}}(X_{i,k}; \mathbf{Y} | \mathbf{U}_k)/N_k \xrightarrow{\text{p}} \mathbb{R}_k. \quad (36)$$

*Proof:* This proof will show (35) only. The proof of (36) is similar. Partition the $N$-bit interleaver $\mathbf{\Pi} = [\pi_1, \cdots, \pi_N]$ into $n$ $m$-bit long blocks as $\mathbf{\Pi} = [\mathbf{\Pi}_1, \cdots, \mathbf{\Pi}_n]$, where $N = nm$ and $\mathbf{\Pi}_j = [\pi_{(j-1)m+1}, \cdots, \pi_{jm}]$ for $j = 1, \cdots, n$. The subsequences that lie within the $j$th block are denoted by $\mathbf{X}_k^{(j)} = \langle \mathbf{X}_k \rangle_{(j-1)m+1}^{jm}$, $\mathbf{Y}_k^{(j)} = \langle \mathbf{Y}_k \rangle_{(j-1)m+1}^{jm}$, and $\mathbf{U}_k^{(j)} = \langle \mathbf{U}_k \rangle_{(j-1)m+1}^{jm}$. The length of $\mathbf{X}_k^{(j)}$ is denoted by $N_k^{(j)}$.

Since the channel is stationary, ergodic, and indecomposable [27] and so is the random interleaver, it can be shown (see [27] and [18]) that as $m \to \infty$

$$\lim_{N \to \infty} I_{\mathbf{\Pi}}(\mathbf{X}_k; \mathbf{Y} | \mathbf{U}_k) = \lim_{m \to \infty} \sum_{j=1}^n I_{\mathbf{\Pi}_j}(\mathbf{X}_k^{(j)}; \mathbf{Y}_k^{(j)} | \mathbf{U}_k^{(j)}).$$

Furthermore, $\lim_{m \to \infty} n N_k^{(j)}/N_k = 1$ for any $j$. Therefore

$$\lim_{N \to \infty} \frac{1}{N_k} I_{\mathbf{\Pi}}(\mathbf{X}_k; \mathbf{Y} | \mathbf{U}_k)$$
$$= \lim_{n \to \infty} \lim_{m \to \infty} \frac{1}{n N_k^{(j)}} \sum_{j=1}^n I_{\mathbf{\Pi}_j}(\mathbf{X}_k^{(j)}; \mathbf{Y}_k^{(j)} | \mathbf{U}_k^{(j)})$$
$$= \lim_{m \to \infty} E_{\mathbf{\Pi}_j} \left[ \frac{1}{N_k^{(j)}} I_{\mathbf{\Pi}_j}(\mathbf{X}_k^{(j)}; \mathbf{Y}_k^{(j)} | \mathbf{U}_k^{(j)}) \right] \quad (37)$$
$$= \mathbb{C}_k^{\text{i.u.d.}} \quad (38)$$

where (37) is due to the stationarity and ergodicity of the random interleaver, and (38) is due to the definition (31) and the fact that $\mathbf{\Pi}_j$ and $\mathbf{\Pi}$ have the same statistics. ∎

Therefore, if a specific weight distribution yields an optimal ensemble average achievable rate, we can generate a sufficiently long interleaver realization to achieve the same rate.

The random interleaver has the asymptotic property similar to that of the rectangular interleaver. If $w_k \to 0$, symbols of codeword $k$ are expected to be scattered far away from each other and $\mathbb{R}_k$ will approach $\mathbb{C}_k^{\text{i.u.d.}}$ as shown in the following lemma.

*Lemma 2:* Let $w_k$ be the weight of subchannel $k$, then

$$\lim_{w_k \to 0} \left( \mathbb{C}_k^{\text{i.u.d.}} - \mathbb{R}_k \right) = 0, \quad k = 1, \cdots, K.$$

*Proof:* By the definitions in (31) and (32) and the chain rule of mutual information, we get

$$\mathbb{C}_k^{\text{i.u.d.}} - \mathbb{R}_k$$
$$= \lim_{N \to \infty} E_{\boldsymbol{\Pi}} \left[ \frac{1}{N_k} I_{\boldsymbol{\Pi}}(\mathbf{X}_k; \mathbf{Y}|\mathbf{U}_k) \right.$$
$$\left. - \frac{1}{N_k} \sum_{i=1}^{N_k} I_{\boldsymbol{\Pi}}(X_{i,k}; \mathbf{Y}|\mathbf{U}_k) \right]$$
$$= \lim_{N \to \infty} E_{\boldsymbol{\Pi}} \left[ \frac{1}{N_k} \sum_{i=1}^{N_k} \left( I_{\boldsymbol{\Pi}}(X_{i,k}; \mathbf{Y}|\mathbf{U}_k, \right.\right.$$
$$\left.\left. X_{1,k}, \cdots, X_{i-1,k}) - I_{\boldsymbol{\Pi}}(X_{i,k}; \mathbf{Y}|\mathbf{U}_k) \right) \right]. \quad (39)$$

Let the distance between the consecutive bits $X_{i,k}$ and $X_{i-1,k}$ be

$$D_{i,k} = t(i,k) - t(i-1,k), \quad i = 2, \cdots, N_k.$$

The random sequence $\{D_{i,k}\}_{i=2}^{N_k}$ is i.i.d. with probability mass function

$$\Pr(D_{i,k} = d) = w_k(1-w_k)^{d-1}, \quad d \geq 1.$$

Let $\alpha = \ln(2)^{-1} \max_{1 \leq i \leq Q, 1 \leq j \leq Q} d(\mathbf{p}_i, \mathbf{p}_j)$. Applying Lemma 1 to (39) yields

$$\mathbb{C}_k^{\text{i.u.d.}} - \mathbb{R}_k < \lim_{N \to \infty} E_{\boldsymbol{\Pi}} \left[ \frac{1}{N_k} \sum_{i=2}^{N_k} \alpha \tau(P)^{D_{i,k}-1} \right]$$
$$= E_{\boldsymbol{\Pi}} \left[ \alpha \tau(P)^{D_{i,k}-1} \right]$$
$$= \sum_{D_{i,k}=1}^{\infty} \alpha w_k (1-w_k)^{D_{i,k}-1} \tau(P)^{D_{i,k}-1}$$
$$= \frac{\alpha w_k}{1 - \tau(P)(1-w_k)}. \quad (40)$$

Since $0 \leq \tau(P) < 1$, letting $w_k \to 0$ in (40) completes the proof. ∎

This implies that a random interleaver with equal weight $w_k = 1/K$ is asymptotically optimal as follows.

*Theorem 2:* For a $K$-level random interleaver with weight distribution $\mathbf{w} = [1/K, \cdots, 1/K]$

$$\lim_{K \to \infty} \mathbb{R} = C^{\text{i.u.d.}}.$$

*Proof:* Since the i.u.d. capacity is independent of any interleaving scheme, we have $C^{\text{i.u.d.}} = \mathbb{C}^{\text{i.u.d.}} = \sum_{k=1}^K w_k \mathbb{C}_k^{\text{i.u.d.}}$. Substitute $w_k = 1/K$ into (40) and use (33) and (34), we get

$$C^{\text{i.u.d.}} - \mathbb{R} = \sum_{k=1}^K K^{-1}(\mathbb{C}_k^{\text{i.u.d.}} - \mathbb{R}_k)$$
$$< \frac{\ln(2)^{-1} \max_{1 \leq i \leq Q, 1 \leq j \leq Q} d(\mathbf{p}_i, \mathbf{p}_j) K^{-1}}{1 - \tau(P)(1 - K^{-1})}.$$

Letting $K \to \infty$ completes the proof. ∎

### B. Optimization of the Random Interleaver

Note that $\mathbb{R}_k(\mathbf{w})$ depends on $\mathbf{w}$ only through $\sum_{i=1}^{k-1} w_i$, the percentage of the training symbols at subchannel $k$. It is useful to quantify the utility (in terms of mutual information) of the randomly positioned pilot symbols.

*Definition 5:* The pilot-utility function $\mu(x)$ is defined as the achievable rate of the data as a function of the pilot percentage $x$

$$\mu(x) = \lim_{n \to \infty} I(X_t; \langle \mathbf{Y} \rangle_{t-n}^{t+n} | \langle \mathbf{Z} \rangle_{t-n}^{t-1}, \langle \mathbf{Z} \rangle_{t+1}^{t+n}) \quad (41)$$

for $x \in [0,1]$ where

$$Z_t = \begin{cases} X_t, & \text{with probability } x \\ \phi, & \text{with probability } 1-x. \end{cases}$$

The function $\mu(x) : [0,1] \longrightarrow [0,1]$ is assumed to be continuously differentiable in $(0,1)$ and is shown to be monotonically increasing $\mu(x) \leq \mu(y)$ for $0 \leq x \leq y \leq 1$ in Appendix II. A Monte-Carlo method similar to that in Section III-B can be used to estimate $\mu(x)$. The pilot-utility function can also be viewed as the EXIT function where the *a priori* probabilities are passed through a binary erasure channel (BEC) [21]. The following two theorems show that $\mathbb{R}_k$ and $\mathbb{C}_k^{\text{i.u.d.}}$ are simply the evaluations on $\mu(x)$.

*Theorem 3:* Let $\mu(x)$ be the pilot-utility function of the FSMC. The average achievable rate of level $k$ in successive decoding under a random interleaver is

$$\mathbb{R}_k(\mathbf{w}) = \mu\left( \sum_{j=1}^{k-1} w_j \right) \quad (42)$$

for $k = 1, \cdots, K$. The overall average achievable rate is

$$\mathbb{R}(\mathbf{w}) = \sum_{k=1}^K w_k \mu\left( \sum_{j=1}^{k-1} w_j \right). \quad (43)$$

*Proof:* As $N \to \infty$, almost all terms $I_{\boldsymbol{\Pi}}(X_{i,k}; \mathbf{Y}|\mathbf{U}_k)$ inside the summation of (32) will converge to one windowed term $\lim_{n \to \infty} I_{\boldsymbol{\Pi}}(X_{i,k}; \langle \mathbf{Y} \rangle_{t(i,k)-n}^{t(i,k)+n} | \langle \mathbf{U}_k \rangle_{t(i,k)-n}^{t(i,k)+n})$ due to the exponential decay of the FSMC channel memory as shown in Lemma 1, also see [18]. So the definition of $\mathbb{R}_k$ in (32) can be re-written as

$$\mathbb{R}_k(\mathbf{w}) = \lim_{n \to \infty} E_{\boldsymbol{\Pi}} \left[ I_{\boldsymbol{\Pi}}(X_{i,k}; \langle \mathbf{Y} \rangle_{t(i,k)-n}^{t(i,k)+n} | \langle \mathbf{U}_k \rangle_{t(i,k)-n}^{t(i,k)+n}) \right]. \quad (44)$$

By the construction of the random interleaver, $\mathbf{U}_k = [U_1, \cdots, U_N]$, where

$$U_t = \begin{cases} X_t, & \text{with probability } \sum_{i=1}^{k-1} w_i \\ \phi, & \text{with probability } 1 - \sum_{i=1}^{k-1} w_i \end{cases}$$

for $t \neq t(i,k)$, is equivalent to a sequence of random training symbols. Therefore, from (41) and (44) and the stationarity and ergodicity of both the channel and the interleaver, we have $\mathbb{R}_k(\mathbf{w}) = \mu(\sum_{j=1}^{k-1} w_j)$. Then (43) holds since $\mathbb{R} = \sum_{k=1}^K w_k \mathbb{R}_k$. ∎

*Theorem 4:* Area property. Let $\mu(x)$ be the pilot-utility function of the FSMC. The average i.u.d. capacity of level $k$ in successive decoding under a random interleaver is

$$\mathbb{C}_k^{\text{i.u.d.}} = \frac{1}{w_k} \int_{\sum_{j=1}^{k-1} w_j}^{\sum_{j=1}^k w_j} \mu(x) \, dx \quad (45)$$

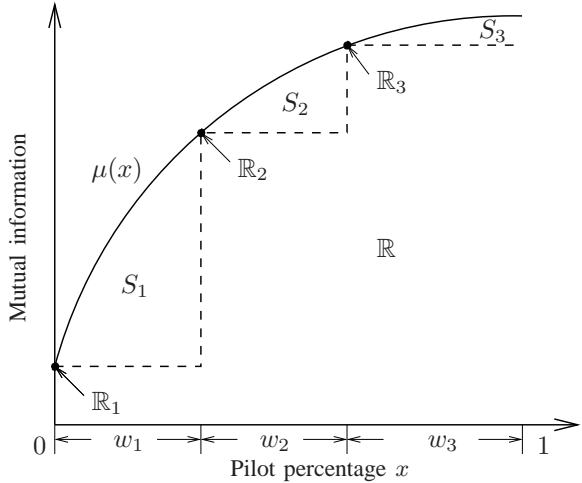

Fig. 1. Illustration of the average achievable rate for a 3-level successive decoding under the random interleaver in relation to the pilot-utility function.

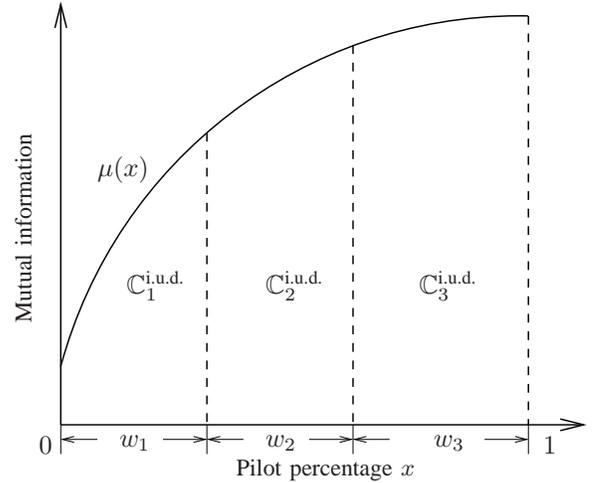

Fig. 2. Illustration of the average i.u.d. capacity for a 3-level successive decoding under the random interleaver in relation to the pilot-utility function.

where $w_k \neq 0$ for $k = 1, \cdots, K$. The overall i.u.d. capacity is

$$C^{\text{i.u.d.}} = \int_0^1 \mu(x)dx. \quad (46)$$

*Proof:* Let $\mathbf{w} = [w_1, \cdots, w_K]$ be the weight distribution of a $K$-level random interleaving scheme. Consider a new interleaving scheme that further divides the subchannel $k$ of weight $w_k$ into $m$ sub-subchannels of weights $w_k/m$. These sub-subchannels are indexed by $k_1, \cdots, k_m$ and the new weight distribution is $\mathbf{w} = [w_1, \cdots, w_{k-1}, w_k/m, \cdots, w_k/m, w_{k+1}, \cdots, w_K]$. Similar to (16), by the chain rule of mutual information we have

$$\mathbb{C}_k^{\text{i.u.d.}} = \sum_{i=1}^m \frac{1}{m} \mathbb{C}_{k_i}^{\text{i.u.d.}}. \quad (47)$$

As $m \to \infty$, $w_k/m \to 0$ and $\mathbb{C}_{k_i}^{\text{i.u.d.}} \to \mathbb{R}_{k_i}$ for $i = 1, \cdots, m$ from Lemma 2. Therefore, let $m \to \infty$ in the right hand side of (47), we have

$$\mathbb{C}_k^{\text{i.u.d.}} = \lim_{m \to \infty} \frac{1}{m} \sum_{i=1}^m \mathbb{R}_{k_i} \quad (48)$$

$$= \lim_{m \to \infty} \frac{1}{m} \sum_{i=1}^m \mu\left(\sum_{j=1}^{k-1} w_j + \frac{w_k}{m}(i-1)\right) \quad (49)$$

$$= \frac{1}{w_k} \int_{\sum_{j=1}^{k-1} w_j}^{\sum_{j=1}^{k} w_j} \mu(x)dx \quad (50)$$

where (48) is due to Lemma 2, (49) is due to (42), and (50) is due to the continuity of $\mu(x)$. Equation (46) then follows (33). ∎

The Theorem 3 and 4 are illustrated graphically in Fig. 1 and 2, respectively. Fig. 1 shows that $\mathbb{R}_k$ is equal to $\mu(x)$ evaluated at $\sum_{j=1}^{k-1} w_j$ and $\mathbb{R}$ is the area under a stair-like curve. Fig. 2 shows that $\mathbb{C}_k^{\text{i.u.d.}}$ is the area beneath $\mu(x)$ between $x = \sum_{j=1}^{k-1} w_j$ and $x = \sum_{j=1}^{k} w_j$ normalized by $w_k$ and $C^{\text{i.u.d.}}$ is the area underneath $\mu(x)$ between $x = 0$ and $x = 1$. The rate loss of doing SED at level $k$ is the normalized area $S_k = \frac{1}{w_k} \int_{\sum_{j=1}^{k-1} w_j}^{\sum_{j=1}^{k} w_j} \mu(x)dx - \mu(\sum_{j=1}^{k-1} w_j)$ between the stair-like curve and the pilot-utility function as shown in Fig. 1. The statement (46) was also shown as the area property of an EXIT function [21].

The optimal weight distribution of a random interleaver is the solution to the following maximization problem:

$$\max_{\mathbf{w}} \mathbb{R}(\mathbf{w}) = \sum_{k=1}^K w_k \mu\left(\sum_{j=1}^{k-1} w_j\right) \quad (51)$$

subject to

$$\sum_{k=1}^K w_k = 1$$

and

$$0 \leq w_k \leq 1, \quad k = 1, \cdots, K.$$

Before solving (51), we show the property of an optimal solution that it is always advantageous to use all levels allowed.

*Proposition 2:* If $\mu(x)$ is strictly increasing, then $\mathbb{R}(\mathbf{w}_{K+1}^*) > \mathbb{R}(\mathbf{w}_K^*)$, where $\mathbf{w}_K^*$ and $\mathbf{w}_{K+1}^*$ are the optimal weight distributions of the $K$-level and $(K+1)$-level successive decoding schemes, respectively.

*Proof:* Let $\mathbf{w}_K^* = [w_1, \cdots, w_K]$ be the optimal point. Without loss of generality, we assume $w_i > 0$ for all $i$ because a $K$-level scheme with a zero-weight level is equivalent to a $(K-1)$-level scheme. Now construct a $(K+1)$-level weight distribution $\mathbf{w}_{K+1} = [w_1', \cdots, w_{K+1}']$ by splitting $\mathbf{w}_K^*$ into two equally weighted levels so that $w_i' = w_i$ for $i = 1, \ldots, K-1$ and $w_i' = w_K/2$ for $i = K, K+1$.

From (42), $\mathbb{R}_i(\mathbf{w}_{K+1}) = \mathbb{R}_i(\mathbf{w}_K^*) = \mu(\sum_{j=1}^{i-1} w_j)$ for $i = 1, \cdots, K$. By the strict monotonicity of $\mu(x)$, we have

$$\mathbb{R}_{K+1}(\mathbf{w}_{K+1}) = \mu\left(\sum_{i=1}^{K-1} w_i' + \frac{w_K}{2}\right) > \mathbb{R}_K(\mathbf{w}_K^*).$$

From (34) and (51), there exists an optimal point $\mathbf{w}_{K+1}^*$ so that

$$\mathbb{R}(\mathbf{w}_{K+1}^*) \geq \mathbb{R}(\mathbf{w}_{K+1}) > \mathbb{R}(\mathbf{w}_K^*).$$

A direct result of Proposition 2 is that an optimal weight distribution must not have zero terms as follows.

*Proposition 3:* The local optimum $\mathbf{w}^*$ satisfies that $w_k^* > 0$ for $k = 1, \cdots, K$ if $\mu(x)$ is strictly increasing.

The Karush-Kuhn-Tucker (KKT) conditions are necessary for the solution of the nonlinear maximization problem (51) with both equality and inequality constraints to be optimal, see for example [35]. Assume $\mathbb{R}(\mathbf{w}) : [0,1]^K \longrightarrow [0,1]$ is continually differentiable. It can be verified that the constant rank constraint qualification (CRCQ) [35] holds for (51), then the local optimum $\mathbf{w}^*$ satisfies that

$$\nabla \mathbb{R}(\mathbf{w}^*) + \sum_{i=1}^{K} \nu_i(-w_i^*) + \lambda\left(1 - \sum_{i=1}^{K} w_i^*\right) = 0 \quad (52)$$

$$\nu_i w_i^* = 0, \quad i = 1, \cdots, K \quad (53)$$

$$w_i^* \geq 0, \quad i = 1, \cdots, K \quad (54)$$

for some $\lambda$ and $\nu_i \geq 0$ for $i = 1, \cdots, K$. We assume that $\mu(x)$ is strictly increasing in the derivation below. From Proposition 3, $w_j^* > 0$, thus $\nu_j = 0$ for $j = 1, \cdots, K$ due to (53). By (43) and (42) and some straightforward manipulation, the necessary conditions (52) to (54) can be simplified to

$$\mu(0) + w_2^* \mu'\left(\sum_{i=1}^{1} w_i^*\right) + w_3^* \mu'\left(\sum_{i=1}^{2} w_i^*\right) + \cdots$$
$$+ w_{K-1}^* \mu'\left(\sum_{i=1}^{K-2} w_i^*\right) + w_K^* \mu'\left(\sum_{i=1}^{K-1} w_i^*\right) = \lambda \quad (55)$$

$$\mu\left(\sum_{i=1}^{1} w_i^*\right) + w_3^* \mu'\left(\sum_{i=1}^{2} w_i^*\right) + \cdots$$
$$+ w_{K-1}^* \mu'\left(\sum_{i=1}^{K-2} w_i^*\right) + w_K^* \mu'\left(\sum_{i=1}^{K-1} w_i^*\right) = \lambda \quad (56)$$

$$\vdots$$

$$\mu\left(\sum_{i=1}^{K-2} w_i^*\right) + w_K^* \mu'\left(\sum_{i=1}^{K-1} w_i^*\right) = \lambda \quad (57)$$

$$\mu\left(\sum_{i=1}^{K-1} w_i^*\right) = \lambda \quad (58)$$

$$\sum_{i=1}^{K} w_i^* = 1 \quad (59)$$

$$w_i > 0, \quad i = 1, \cdots, K \quad (60)$$

where $\mu'$ denotes the first order derivative of $\mu$.

Define $\sigma_k = \sum_{i=1}^{k-1} w_i$ for $k = 1, \cdots, K$. Since $\mu(x)$ is continuous and strictly increasing, the inverse $\mu^{-1}(y)$ exists, where $\mu(0) \leq y \leq \mu(1)$. From (59) and (58), for any $\mu(0) \leq \lambda \leq \mu(1)$, we find $w_K$ as

$$\sigma_K = \mu^{-1}(\lambda) \quad (61)$$

$$w_K = \max(1 - \sigma_K, 0) \quad (62)$$

and find $w_{i-1}$ for $i = K, \cdots, 2$ recursively as

$$\sigma_{i-1} = \mu^{-1}(\mu(\sigma_i) - w_i \mu'(\sigma_i))$$
$$w_{i-1} = \max(\sigma_i - \sigma_{i-1}, 0). \quad (63)$$

The set of local optimal points $\{\mathbf{w}^*\}$ is given by $\lambda$ that satisfies

$$\theta(\lambda) = 1 - \sum_{i=1}^{K} w_i = 0, \quad \mu(0) \leq \lambda \leq \mu(1).$$

The global optimal solution is $\arg\max_{\mathbf{w} \in \{\mathbf{w}^*\}} \mathbb{R}(\mathbf{w})$.

For many channels, inserting more pilot symbols has diminishing return in the achievable rate of the data. The pilot-utility functions of these channels are concave and the optimal strategy is to allocate more weight at higher levels.

*Proposition 4:* If $\mu(x) : [0,1] \longrightarrow [0,1]$ is continually differentiable in $(0,1)$, strictly increasing, and concave, the optimal weight distribution satisfies $w_1^* \leq w_2^* \leq \cdots \leq w_K^*$.

*Proof:* From (55) to (58), we have

$$\mu\left(\sum_{i=1}^{k-1} w_i^*\right) + w_{k+1}^* \mu'\left(\sum_{i=1}^{k} w_i^*\right) = \mu\left(\sum_{i=1}^{k} w_i^*\right) \quad (64)$$

for $k = 1, \cdots, K-1$. From the concavity of $\mu(x)$, we have

$$\mu\left(\sum_{i=1}^{k} w_i^*\right) \geq \mu\left(\sum_{i=1}^{k-1} w_i^*\right) + w_k^* \mu'\left(\sum_{i=1}^{k} w_i^*\right) \quad (65)$$

for $k = 1, \cdots, K-1$. The combination of (64), (65), and the fact that $\mu'(x) > 0$ shows that

$$w_{k+1}^* \geq w_k^*, \quad k = 1, \cdots, K-1.$$

∎

### C. A Construction of the Deterministic Interleaver

Unlike random interleavers, the optimization of deterministic interleavers has combinatorial complexity. Although the optimal placement of pilot symbols for PSAM has been studied in the literature [36], the problem here is more difficult as there are $K$ codewords to be placed. Thus, we present a family of deterministic interleavers that are constructed from empirical rules proposed for the Markov fading channel.

First, more weight shall be allocated to higher levels because of Proposition 4. Second, the weight of level 1 (pilot percentage) shall be optimized as it has zero achievable rate and the most mutual information loss. Optimization with respect to $w_1$ often yields the most gain. Third, it is desirable to separate the symbols within a codeword and to place the symbols from lower levels evenly around them. Accordingly, we construct a family of binary-weighted interleavers with $w_{k+1} = 2w_k$ for $k \geq 2$ as

$$\boldsymbol{\pi} = [\boldsymbol{\omega}, \cdots, \boldsymbol{\omega}], \quad \boldsymbol{\omega} = [1, \boldsymbol{v}_K, \cdots, \boldsymbol{v}_K] \in \mathbb{K}^L. \quad (66)$$

Here the vector $\boldsymbol{v}_K$ is defined recursively as

$$\boldsymbol{v}_K = [K, v_{K-1}(1), K, v_{K-1}(2), \cdots,$$
$$K, v_{K-1}(2^{K-2} - 1), K]$$

for $K > 2$ and $\boldsymbol{v}_2 = [2]$. For example, a 3-level binary-weight interleaver has $\boldsymbol{v}_3 = [3, 2, 3]$, a 4-level one has $\boldsymbol{v}_4 = [4, 3, 4, 2, 4, 3, 4]$, and a 5-level one has $\boldsymbol{v}_5 = [5, 4, 5, 3, 5, 4, 5, 2, 5, 4, 5, 3, 5, 4, 5]$. It is clear that the training symbols are well placed for each level. Furthermore, the weight of level 1 can be optimized by finding the optimal number of $\boldsymbol{v}_K$ in (66).

## V. DESIGN FOR ITERATIVE ESTIMATION AND DECODING

For the iterative receivers, the design techniques proposed in Section IV including interleaver optimization and code rate allocation are no longer optimal because the achievable rate becomes too conservative a performance measure. On the other hand, the EXIT chart based analysis [26] was shown to predict the convergence behavior of an iterative process very well. Hence, given a family of codes of various rates we use the EXIT chart to find the maximal code rate supported by an IED algorithm at each subchannel and to formulate the interleaver optimization problem.

### A. EXIT Function of the Estimator

Let $\{L^{e,in}(x_{i,k})\}$ and $\{L^{e,out}(x_{i,k})\}$ in (13) be the sequence of likelihood ratios (extrinsic information) at the input and the output of the estimator, respectively. They are assumed to be the realizations of i.i.d. random variables. The input and output mutual information for the estimator can be obtained from

$$I_k^{e,in} = I\left(X_{i,k}; L^{e,in}(X_{i,k})\right)$$
$$= \lim_{N \to \infty} \left(1 - \frac{1}{N_k} \sum_{i=1}^{N_k} E\left[-\log_2 \frac{L^{e,in}(x_{i,k})}{1+L^{e,in}(x_{i,k})}\right]\right) \quad (67)$$

and

$$I_k^{e,out} = I\left(X_{i,k}; L^{e,out}(X_{i,k})\right)$$
$$= \lim_{N \to \infty} \left(1 - \frac{1}{N_k} \sum_{i=1}^{N_k} E\left[-\log_2 \frac{L^{e,out}(x_{i,k})}{1+L^{e,out}(x_{i,k})}\right]\right). \quad (68)$$

For a given channel and interleaver, the estimator EXIT function at level $k$ is

$$I_k^{e,out} = T_k(I_k^{e,in}). \quad (69)$$

In order to estimate (69), a sequence $\{L^{e,in}(x_{i,k})\}$, the information content of which is measured according to (67), is generated according to a given PDF with a single parameter and fed to the estimator. The estimator output $\{L^{e,out}(x_{i,k})\}$ is then collected to produce an estimate of the output mutual information using (68). The entire curve of $I_k^{e,out} = T_k(I_k^{e,in})$ can be traced by varying the single parameter of the PDF so that $I_k^{e,in}$ changes from 0 to 1.

The exact PDF of $L^{e,in}(x_{i,k})$ is difficult to obtain. One commonly adopted approach is to assume that $L^{e,in}(x_{i,k})$ is derived from an AWGN channel $Y = X + W$ with noise variance $E[W^2] = \sigma_w^2$ so that $L^{e,in}(x_{i,k}) \sim \mathcal{N}(2/\sigma_w^2, 4/\sigma_w^2)$. This approach will be used for deterministic interleavers. However, for random interleavers, (69) can be computed more efficiently using the pilot-utility function (41). Let $L^{e,in}(x_{i,k})$ be drawn according to the following distribution

$$L^{e,in}(x_{i,k}) = \begin{cases} +\infty, & \text{with probability } x \\ 1, & \text{with probability } 1-x \end{cases}$$

which means that a symbol is completely known with probability $x$. Thus, the input mutual information is

$$I_k^{e,in} = I(X_{i,k}; L^{e,in}(X_{i,k})) = x$$

and, by the definition of $\mu(x)$ in (41), the output mutual information is

$$T_k(x) = \mu\left(xw_k + \sum_{i=1}^{k-1} w_i\right), \quad x \in [0,1] \quad (70)$$

where $\mathbf{w} = [w_1, \cdots, w_K]$ is the weight distribution.

### B. EXIT Function of the Decoder

Let $\mathcal{C}(r)$ be a code of rate $r$. Let the EXIT function of its decoder be

$$I_k^{d,out} = T_d(I_k^{d,in}, \mathcal{C}(r)).$$

Since the FSMC considered here is a fading channel with good channel estimation at the receiver. It is convenient to assume that the decoder input extrinsic information $L^{e,in}(x_{i,k})$ is derived from a known-state fading channel $Y = HX + W$, where $H \sim \mathcal{CN}(0,1)$ and $W \sim \mathcal{CN}(0, \sigma_w^2)$. Then $I_k^{d,in}$ is a function of the AWGN variance only and is equal to the i.u.d. binary-input capacity of a known-state fading channel given in [18]

$$I_k^{d,in} = \frac{\lambda_2 F(\lambda_1+1, 1; \lambda_1+2; -1)}{(\lambda_1+\lambda_2)(\lambda_1+1)\ln 2} - \frac{\lambda_1 F(\lambda_2, 1; \lambda_2+1; -1)}{\lambda_2(\lambda_1+\lambda_2)\ln 2} \quad (71)$$

where $\lambda_{1,2} = \frac{1}{2}(\sqrt{1+\sigma_w^2} \mp 1)$ and $F(a,b;c;z)$ is a hypergeometric function, or a Gauss hypergeometric function. The output mutual information $I_k^{d,out}$ can be measured at the soft output of the decoder.

### C. Design Using the EXIT Charts

The EXIT chart is a diagram where the estimator EXIT function $I_k^{e,out} = T_k(I_k^{e,in})$ and the inverse of the decoder EXIT function $I^{d,in} = T_d^{-1}(I^{d,out}, \mathcal{C}(r))$ are plotted together. The iterative process can be tracked on the EXIT chart as a flow of mutual information with initial value $I_k^{e,in} = 0$. As long as $T_k(x) > T_d^{-1}(x, \mathcal{C}(r))$ for $0 \le x \le 1$, the iteration will proceed to $I^{d,out} = 1$. Hence, the maximal code rate supported by the iterative estimation and decoding at a subchannel can be estimated by

$$r_k^* = \sup_r \{r : T_k(x) - T_d^{-1}(x, \mathcal{C}(r)) > d_t, \ 0 \le x \le 1\} \quad (72)$$

where $d_t \ge 0$ is a design parameter that specifies the allowed minimal tunnel width between two EXIT curves. The code rate at level $k$ is then chosen to be $r_k^*$.

Therefore, we can maximize the overall code rate by matching the estimator EXIT function at each level to the code EXIT function through interleaver design. For random interleavers, it is the following weight distribution optimization problem:

$$\max_{\mathbf{w}} \sum_{k=1}^{K} w_k r_k^* \quad (73)$$

subject to

$$\sum_{k=1}^{K} w_k = 1$$

and

$$0 \le w_k \le 1, \quad k = 1, \cdots, K$$

where $r_k^*$ is given in (72). Note that it is implicitly assumed in (73) that the code EXIT function does not depend on $w_k$ or equivalently $N_k$. The justification is that in the case of $N_k \to \infty$, the dependency of the code EXIT function on $N_k$ becomes rather weak.

## VI. PERFORMANCE ANALYSIS UNDER THE FINITE-LENGTH CONSTRAINT

If the overall delay $N$ of a successive decoding scheme is finite, there is a tradeoff between the number of levels and the codeword length of each level. Clearly, for an infinite $N$ it is always beneficial to increase $K$, while for a very small $N$, the best strategy is to use no more than one code with some training symbols, as observed in [37]. This section provides an analysis of the finite-length effect on the SED based schemes by relating the word-error probability to the codeword length using the random-coding bound [27].

Let $X$ and $Y$, respectively, be the input and output of a memoryless channel and $\Pr(Y|X)$ be the channel transition PDF. The results in [27] state that the error probability of maximum-likelihood decoding of a length-$N$ block code of rate $r$ is upper bounded by

$$\overline{P}_e = 2^{-NE^r(r)} \quad (74)$$

where

$$E^r(r) = \max_{0 \le \rho \le 1} (E^0(\rho) - \rho r) \quad (75)$$

is the random-coding error exponent and

$$E^0(\rho) = -\log_2 \int_Y \Big(\sum_{x=\pm 1} \Pr(x) \Pr(y|x)^{\frac{1}{1+\rho}}\Big)^{1+\rho} dy. \quad (76)$$

Consider a successive decoding scheme using SED and a fixed interleaver $\boldsymbol{\pi}$. Under the SED rule, the $k$th subchannel is treated as a memoryless channel and the probabilities $\{\Pr(\mathbf{Y}|X_{i,k}=a,\mathbf{U}_k)\}_{i=1}^{N_k}$ are assumed to be independent. According to (75) and (76), the random-coding error exponent at the subchannel $k$ for $k=1,\cdots,K$ is

$$E_k^r(r) = \max_{0 \le \rho \le 1} (E_k^0(\rho, \boldsymbol{\pi}) - \rho r) \quad (77)$$

where

$$\begin{aligned} E_k^0(\rho, \boldsymbol{\pi}) &= \lim_{N \to \infty} -\frac{1}{N_k} \sum_{i=1}^{N_k} \log_2 \int_{\mathbf{Y}} \Big(\sum_{a=\pm 1} \Pr(X_{i,k}=a) \\ &\quad \Pr(\mathbf{Y}|X_{i,k}=a,\mathbf{U}_k)^{\frac{1}{1+\rho}}\Big)^{1+\rho} d\mathbf{Y} \\ &= \lim_{N \to \infty} -\frac{1}{N_k} \sum_{i=1}^{N_k} \log_2 \Big\{ \\ &\quad 2^{-\rho} E_{\mathbf{Y}} \Big[\Big(\sum_{a=\pm 1} \Pr(X_{i,k}=a|\mathbf{Y},\mathbf{U}_k)^{\frac{1}{1+\rho}}\Big)^{1+\rho}\Big]\Big\}. \end{aligned} \quad (78)$$

For random interleavers, we take expectation over $\boldsymbol{\Pi}$ to obtain

$$E_k^r(r) = \max_{0 \le \rho \le 1} (E_k^0(\rho) - \rho r) \quad (79)$$

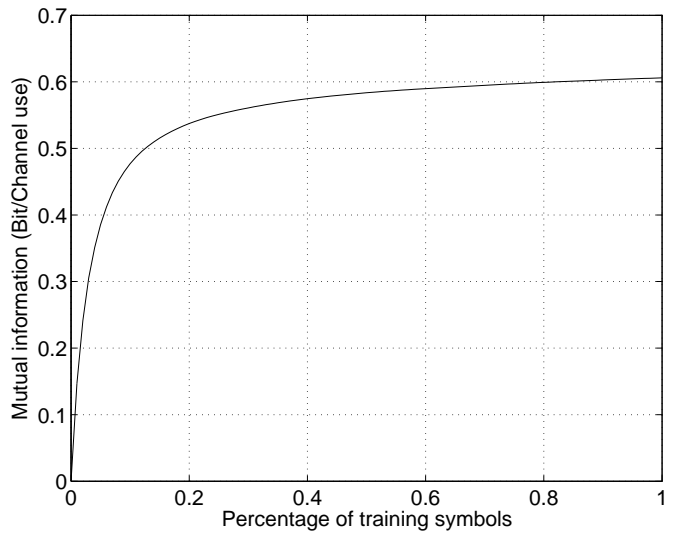

Fig. 3. The pilot-utility function for the example FSMC at $E_s/N_0 = 3$ dB.

where

$$E_k^0(\rho) = E_{\boldsymbol{\Pi}}[E_k^0(\rho, \boldsymbol{\pi})]. \quad (80)$$

Let the word-error probability of all levels be upper bounded by $\overline{P}_e'$. Assume the decoder at each level produces independent errors, the overall error probability of a $K$-level system is upper bounded by $\overline{P}_e = 1 - (1 - \overline{P}_e')^K \le K\overline{P}_e'$. Therefore, for a given total length $N$, a specific interleaver, and a target word-error probability upper bound $\overline{P}_e$, using (74) we can find the upper bound of rate $\overline{r}_k$ at level $k$ by solving

$$E_k^r(\overline{r}_k) = -\frac{1}{N_k} \log_2 \frac{\overline{P}_e}{K} \quad (81)$$

for $\overline{r}_k$. Here $E_k^r(r)$ can be evaluated numerically from (77) and (78) for the deterministic interleaver and from (79) and (80) for the random interleaver. The optimal number of levels under the overall-delay constraints can be found by maximizing the upper bound of the overall rate $\overline{r} = \sum_{k=1}^{K} w_k \overline{r}_k$.

## VII. NUMERICAL RESULTS

### A. Example Channel

A first-order Gauss-Markov process $\widetilde{h}_t = \alpha \widetilde{h}_{t-1} + \sqrt{\alpha^2 - 1} z_t$ is used as the underlying physical channel to derive the finite-state Markov process, where $z_t \sim \mathcal{CN}(0,1)$ is the driving white Gaussian process, $\alpha \in (0,1)$ determines the fading speed, and $\widetilde{h}_t \sim \mathcal{CN}(0,1)$ is the continuous-valued complex channel gain. The channel state space $\{A_1, \cdots, A_Q\}$ is obtained by independently quantizing the real and imaginary part of $\widetilde{h}_t \sim \mathcal{CN}(0,1)$ using the Max-Lloyd algorithm. The state-transition probability is found by integrating the joint PDF of $\widetilde{h}_t$ and $\widetilde{h}_{t+1}$, and the stationary probability by integrating the PDF of $\widetilde{h}_t$. This paper uses the example channel given by $\alpha = 0.95$ and $Q = 36$. The quantization points and boundaries are respectively $\{\pm 1.339, \pm 0.707, \pm 0.225\}$ and $\{\pm 6, \pm 1.023, \pm 0.466, 0\}$ for both dimensions. The FSMC considered here models rather accurately a flat-fading channel with both random phase rotation and magnitude fluctuation.

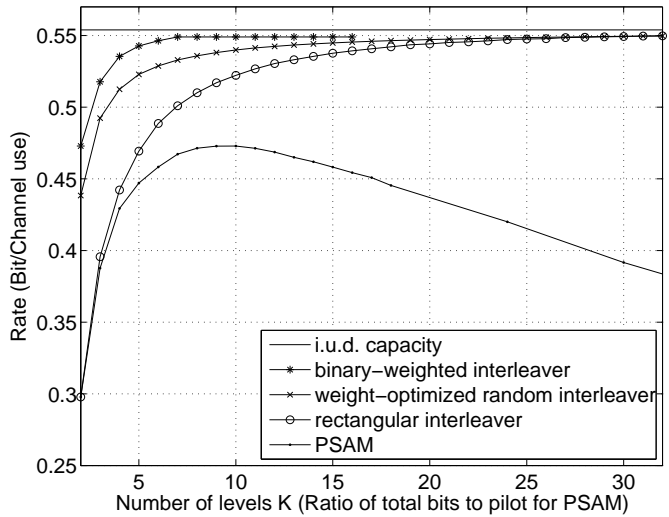

Fig. 4. The achievable rates for various interleavers under the separate estimation and decoding.

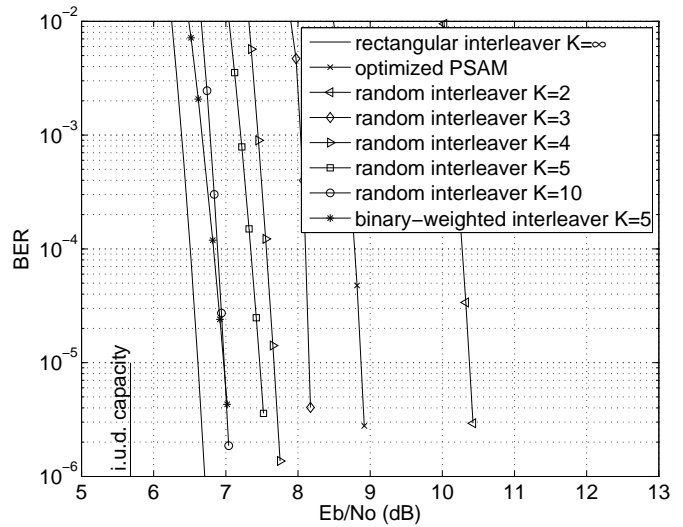

Fig. 5. Error performance comparison of various coding schemes.

## B. Component Codes

The component codes for each level are drawn from a set of irregular LDPC codes optimized for AWGN channels with rates from $0.1$ to $0.7$ with a step of $0.01$. Their degree polynomials are generated by LDPCopt [25]. The decoder uses the message-passing algorithm with 20 iterations. The codeword length is chosen to be proportional to the weight of each level and is sufficiently long. For SED, the code rate is chosen to be the achievable rate i.e. $r_k = R_k$ for the deterministic interleaver and $r_k = \mathbb{R}_k$ for the random interleaver. For IED, the code rates are chosen according to (72), $r_k = r_k^*$, where the tunnel width $d_t = 0$. These rates are then rounded to the nearest available code rates. Note, in some cases, the first few levels have a small codeword length. We will lower the code rates appropriately, usually $0.01$ to $0.03$, to compensate for it.

## C. Design Results for Separate Estimation and Decoding

This section presents design examples of successive decoding with SED. Both the achievable rates and the BERs of actual coding implementation show that the proposed random and binary-weighted interleaver have significant performance gain over the traditional rectangular interleavers and the PSAM.

The pilot-utility function $\mu(x)$ is estimated for the above channel at $E_s/N_0 = 3$ dB as plotted in Fig. 3. Base on $\mu(x)$, we obtain the weight distributions of random interleavers for $K = 2, \cdots, 32$ by solving equations (52) to (54). We also find that the optimal repetition of $\bm{v}_K$ for the binary-weighted interleaver (66) is 9, 5, 3, and 2, respectively, for $K = 2, 3, 4, 5$ and is 1 for $K \geq 6$. Their achievable rates are plotted in Fig. 4. For comparison, Fig. 4 also shows the i.u.d. capacity, the achievable rates of a $K$-level rectangular interleaver and the PSAM. The PSAM is configured to have 1 pilot symbol for every $K-1$ data symbols and in this case the $x$-axis of Fig. 4 is the ratio of the total number of symbols to the number of pilot symbols.

TABLE I
CODE RATES AND LENGTHS FOR RANDOM INTERLEAVERS IN THE SED SCHEME.

| Level | Overall rate | Code rate and length at individual levels | | | | |
|---|---|---|---|---|---|---|
| K=2 | 0.5590 | 0<br>35K | 0.65<br>215K | | | |
| K=3 | 0.5529 | 0<br>21K | 0.48<br>66K | 0.63<br>213K | | |
| K=4 | 0.5556 | 0<br>20K | 0.37<br>50K | 0.56<br>130K | 0.62<br>300K | |
| K=5 | 0.5559 | 0<br>32K | 0.28<br>63K | 0.5<br>130K | 0.59<br>270K | 0.63<br>505K |
| K=10 | 0.5573 | 0<br>21K | 0.16<br>29K | 0.31<br>40.5K | 0.40<br>58.5K | 0.49<br>85.5K |
| | | 0.54<br>123K | 0.56<br>181.5K | 0.59<br>244.5K | 0.61<br>337.5K | 0.62<br>379.5K |

As shown in Fig. 4, the fundamental problem of the PSAM is that the pilot symbols useful for state estimation reduce the overall rate. The successive decoding resolves this problem. The achievable rates of all types of the interleavers considered here are shown to approach $C^{\text{i.u.d.}}$ exponentially fast as $K$ increases. At small $K$ the optimized random interleaver and the binary-weighted interleaver have significant performance gain over the rectangular one. For comparison, at $K = 3$, the rectangular, the random, and the binary-weighted interleaver achieve, respectively, $79.8\%$, $88.9\%$, and $93.5\%$ of the i.u.d. capacity. In order to achieve $95\%$ of the i.u.d. capacity, they would require 11, 6, and 4 levels, respectively. This illustrates the effectiveness of the proposed design for finite $K$. It shall be noted that although the binary-weighted interleaver is shown to outperform the optimized random interleaver in Fig. 4, this result may vary for a different channel because the random interleaver has more degree of freedom for optimization.

For a fair comparison in the code simulation, the target overall rate of all schemes is set to $0.56$. The weight-distributions for the random interleavers are optimized for $E_s/N_0 = 3$ dB. The code rates and lengths are shown in Table I and II for the random interleaver and the binary-weighted interleaver, respectively, in the ascending order of $k = 1, \cdots, K$ from left to right in each row. Note that both the code rate and length increase with the level $k$. The deep rectangular interleaver serves as a benchmark and is designed according to [18] with sufficiently many levels and each level uses the same code of rate 0.56 and length 200K. An optimized PSAM, according to Fig. 4, with 1 pilot for every 9 data symbols and a code of rate 0.62 is also considered.

TABLE II
CODE RATES AND LENGTHS FOR BINARY-WEIGHTED INTERLEAVERS IN THE SED SCHEME.

| Level | Overall rate | Code rate and length at individual levels | | | | |
|---|---|---|---|---|---|---|
| K=5 | 0.5590 | 0 | 0.4 | 0.52 | 0.59 | 0.61 |
| | | 20K | 40K | 80K | 160K | 320K |

TABLE III
CODE RATES AND LENGTHS FOR DIFFERENT INTERLEAVERS IN THE IED SCHEMES.

| Interleavers | Overall rate | Code rate and length at individual levels | | | | |
|---|---|---|---|---|---|---|
| random | 0.5108 | 0 | 0.22 | 0.41 | 0.51 | 0.57 |
| | | 10K | 20K | 55K | 135K | 280K |
| bin-weight | 0.5200 | 0 | 0.37 | 0.49 | 0.54 | 0.57 |
| | | 20K | 40K | 80K | 160K | 320K |
| rectangular | 0.4440 | 0 | 0.54 | 0.55 | 0.56 | 0.57 |
| | | 200K | 200K | 200K | 200K | 200K |
| PSAM | 0.4680 | 0 | 0.52 | | | |
| | | 20K | 180K | | | |

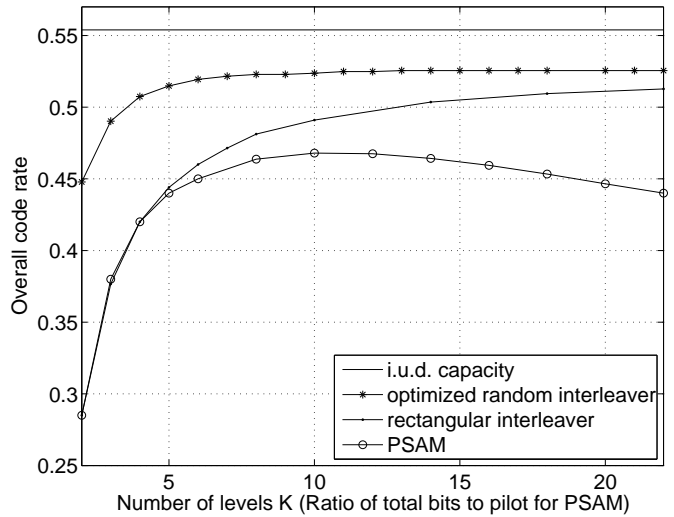

Fig. 6. The overall LDPC code rates supported by the iterative estimation and decoding receiver at $E_s/N_0 = 3$ dB.

Their BER performance is shown in Fig. 5. Both the 5-level binary-weighted interleaver and the 10-level random interleaver perform very close (within 0.3 dB) to the asymptotic deep rectangular interleaver and are around 1.3 dB to the i.u.d. capacity. They have a gain of round 2 dB over PSAM. Note, a 2-level random interleaver performs worse than the PSAM. These results match the achievable rate results in Fig. 4 very well, confirming the good performance of both the random interleaver and the binary-weighted interleaver.

### D. Design Results for Iterative Estimation and Decoding

In the following, we design the IED schemes for various interleavers to maximize the communication rate at a target $E_s/N_0$ of 3 dB. To obtain the weight distribution of the random interleavers, we first estimate the EXIT functions $I^{d,out} = T_d(I^{d,in}, \mathcal{C}(r))$ of the given set of irregular LDPC codes with rates $r = 0.1, 0.11, \cdots, 0.7$ and length 100K. The estimator EXIT functions $T_k$ for $k = 1, \cdots, K$ are given by (70). We then numerically solve (73), where the tunnel width in (72) is set to $d_t = 0$, using the Matlab fmincon function for 200 random initial values. For other schemes, the estimator EXIT charts are computed based on the Gaussian approximated log-likelihood ratio. The resulting overall code rates are plotted in Fig. 6. It shows that the successive decoding with a random interleaver greatly outperforms the rectangular interleaver and PSAM. Compared with SED in Fig. 4, the rate loss of using a small $K$ is much smaller here. In fact, a random interleaver with only 6 levels has a rate very close to its asymptotic case.

The codes for various 5-level successive decoding schemes and the optimized PSAM are specified in Table III. The set of EXIT charts used for code rate selection and optimization is shown in Fig. 7 to 10. For the binary-weighted interleaver, the EXIT functions of the estimator and the decoder match very well, predicting its best performance among all schemes. For the random interleaver, the EXIT chart matches better as the level gets higher, especially at the highest two levels. This explains the good performance for random interleavers. On the other hand, the rectangular interleaver has rather flat estimator EXIT functions and thus little IED gain. Fig. 11 to 14 show the coding results of the above designs in reference to the SNR at which the i.u.d. capacity is equal to the overall rate. All schemes have BER of $10^{-5}$ at around 3.3 dB. Both the random and the binary-weighted interleavers have around 0.6 dB gain from IED, and are, respectively, around 1.1 dB and 1 dB to the i.u.d. capacity. The PSAM also has around 0.6 dB gain for iterative receivers, however, it is around 2 dB to the i.u.d. capacity due to the rate loss of 10% pilot symbols. The performance of a 5-level rectangular interleaver is 2.5 dB to the i.u.d. capacity.

### E. Results for Finite-Length Analysis

Here we show the performance bound of successive decoding with a delay constraint using the random-coding error exponent analysis in Section VI. Both the rectangular and the random interleaver are considered. We compute the error exponents for the rectangular and the random interleaver according to (77) and (79), respectively, using the Monte-Carlo simulation. The overall rate $\overline{r} = \sum_{k=1}^{K} w_k \overline{r}_k$, where $\overline{r}_k$ is computed from (81) with $\overline{P}_e = 10^{-3}$, is shown in Fig. 15 and 16 for the rectangular and the random interleaver, respectively.

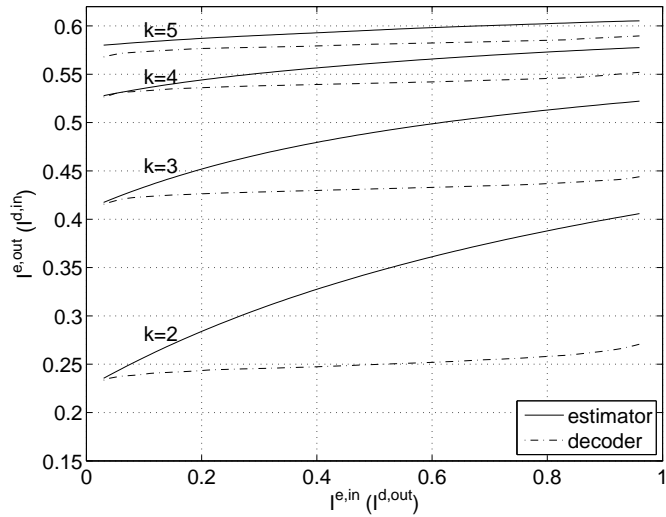

Fig. 7. EXIT charts of the optimized random interleaver.

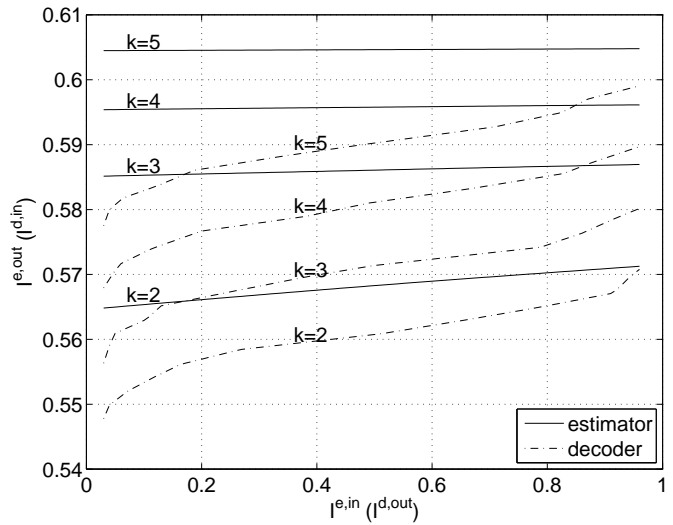

Fig. 9. EXIT charts of the rectangular interleaver.

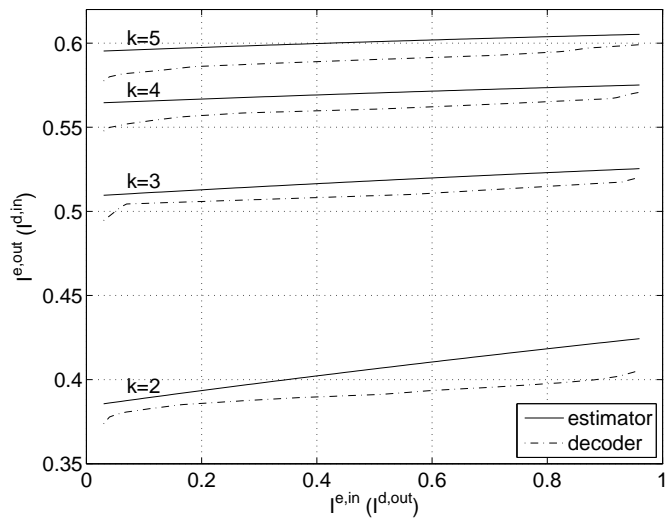

Fig. 8. EXIT charts of the binary-weighted interleaver.

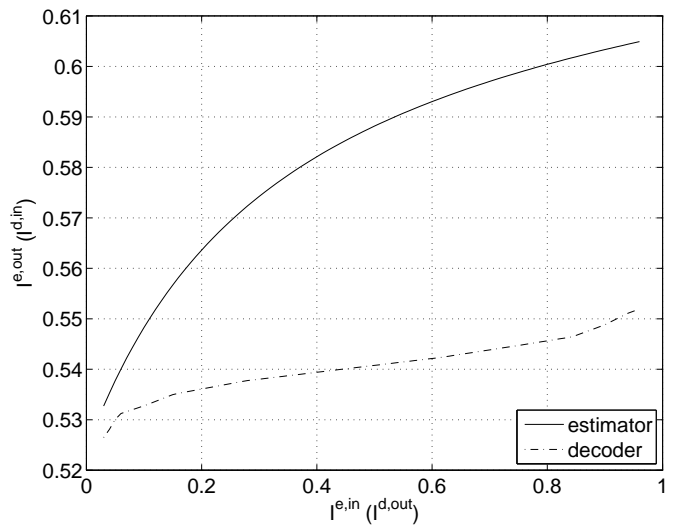

Fig. 10. EXIT charts of PSAM.

In the extreme case of $N = 100$, it is clearly the best to use only one code and some pilot symbols. Otherwise, it is usually beneficial to use more than two levels and there is an optimal number of levels for short to moderate block lengths for both types of interleavers. The random interleaver is shown to have a higher overall rate than the rectangular interleaver especially for small $N$.

## VIII. CONCLUDING REMARKS

In this paper, we have proposed and analyzed new designs of successive decoding scheme with finite levels. The main techniques are a flexible interleaver structure and iterative estimation and decoding within each level. Both the random and the binary-weighted interleaver are constructed to have near i.u.d. capacity performance with as few as five levels. Using irregular LDPC codes, an optimized 10-level random interleaver using SED performs very close to the deep rectangular interleaver, and a 5-level random interleaver using IED is less than 1.1 dB away from the i.u.d. capacity. These results show that successive decoding is not only asymptotically optimal but also attractive for practical systems. The proposed random interleaver also provides some interesting insight into the channel mutual information. We have also analyzed the performance of successive decoding under an overall-delay constraint based on the random-coding error exponent. We showed that using multiple levels is useful for a moderate delay constraint and an optimal number of levels can be found.

The interleaved $K$ LDPC codes used here can be viewed as a compound LDPC code and the successive decoding as a special message-passing schedule. Thus, we have in effect obtained a procedure to construct an irregular LDPC code that can approach the i.u.d. capacity of a realistic channel with memory from a set of AWGN optimized degree profiles. This may suggest a new approach to designing good degree profiles for LDPC codes over channels with memory using the idea of successive decoding.

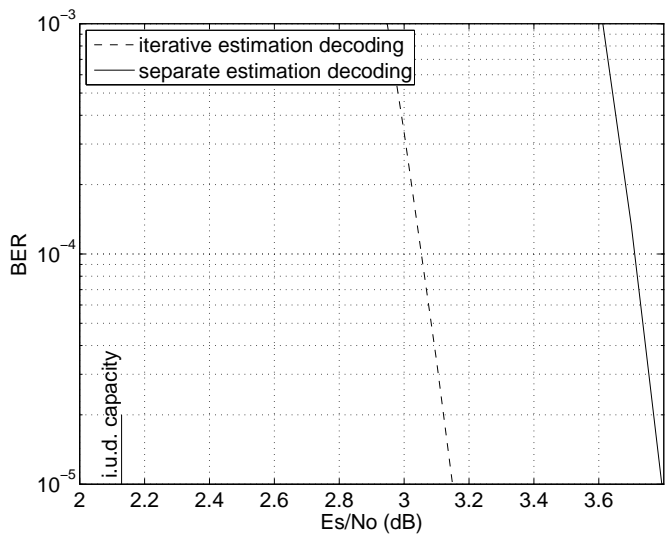

Fig. 11. BER of weight-optimized random interleaver.

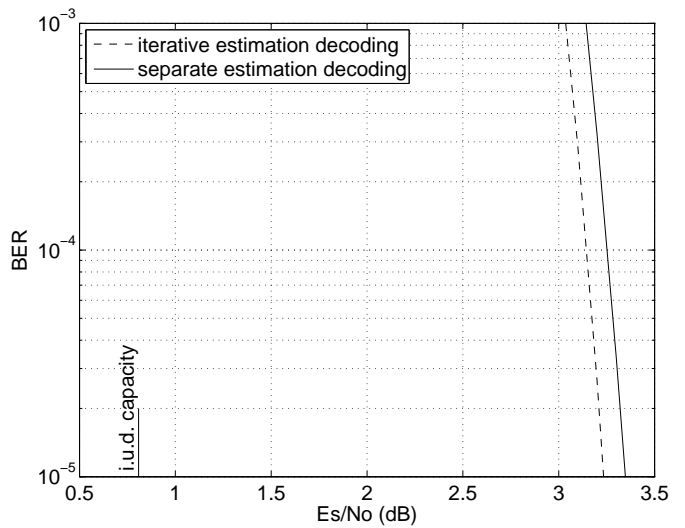

Fig. 13. BER of rectangular interleaver.

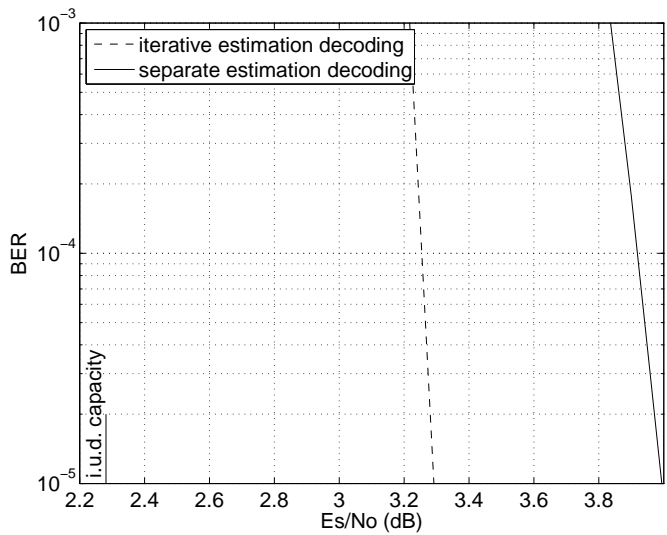

Fig. 12. BER of binary-weighted interleaver.

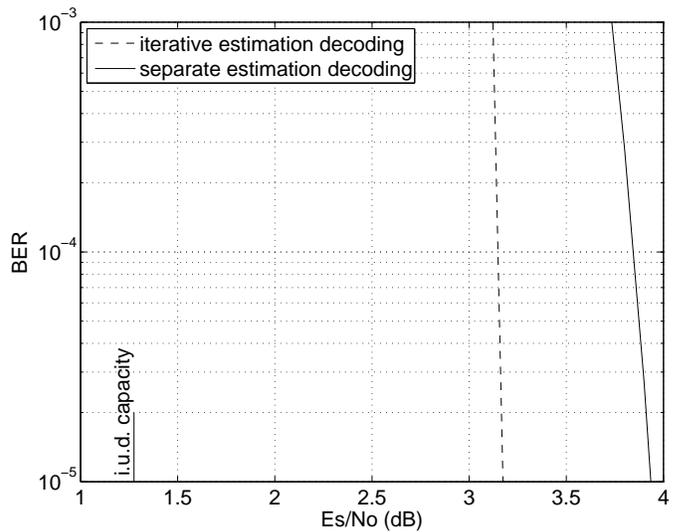

Fig. 14. BER of PSAM.

One possible extension of the current scheme, especially for systems with a stringent delay constraint, is to allow the $K$ decoders to exchange soft information similar to [15] and [16]. Although it is suggested in [15] and [16] that the component codes may be designed based on the original system with perfect decision feedback, more performance gain can be expected if we can find new code-optimization techniques that take into account the iteration between different levels. In practice, the joint construction of the short code and the interleaver under a small delay constraint may further improve the performance.

## APPENDIX I
## PROOF OF LEMMA 1

Let $\Lambda(x_t)$ and $\Lambda'(x_t)$, respectively, be the likelihood ratio of $X_t$ computed using forward recursion windows of $m$ and $m'$ and a backward recursion window of $n$. By the definition of the mutual information

$$I\bigl(X_t; \langle \mathbf{Y}\rangle_{t-m'}^{t+n}|\langle \mathbf{U}\rangle_{t-m'}^{t+n}\bigr) - I\bigl(X_t; \langle \mathbf{Y}\rangle_{t-m}^{t+n}|\langle \mathbf{U}\rangle_{t-m}^{t+n}\bigr) \\ = \frac{1}{\ln 2} E\left[\ln\left(\frac{\Lambda'(x_t)}{1+\Lambda'(x_t)}\frac{1+\Lambda(x_t)}{\Lambda(x_t)}\right)\right]. \quad (82)$$

Let $\alpha_t(q)$ and $\alpha'_t(q)$ be the forward state probabilities computed from windows of $m$ and $m'$, respectively. Let $\beta_t(q)$ be the backward state probability computed from a window of $n$.

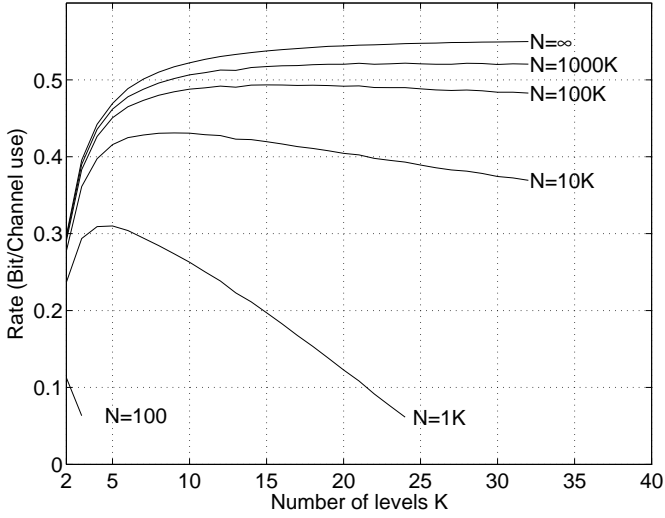

Fig. 15. Performance of rectangular interleavers with finite-length constraint.

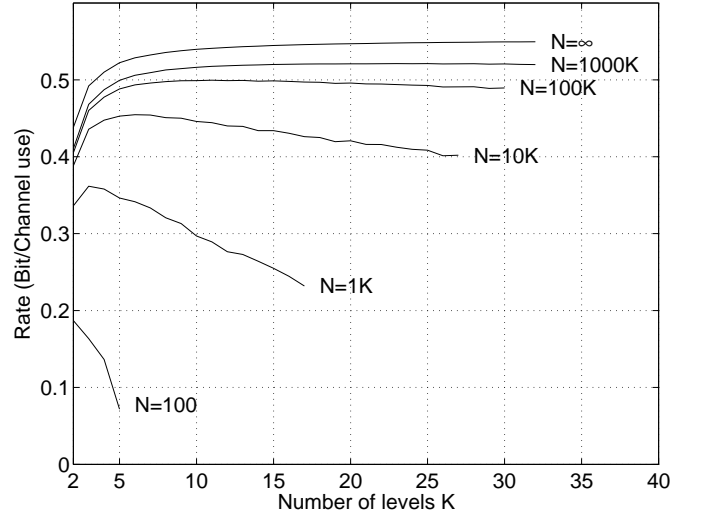

Fig. 16. Performance of random interleavers with finite-length constraint.

By (11), we have

$$\frac{\Lambda'(x_t)}{1+\Lambda'(x_t)}\frac{1+\Lambda(x_t)}{\Lambda(x_t)} = \frac{\sum_{q=1}^{Q}\sum_{q'=1}^{Q}\alpha_t'(q)\gamma_t(q',q,x_t)\beta_{t+1}(q')}{\sum_{q=1}^{Q}\sum_{q'=1}^{Q}\alpha_t(q)\gamma_t(q',q,x_t)\beta_{t+1}(q')}$$

$$\times \frac{\sum_{q=1}^{Q}\sum_{q'=1}^{Q}\alpha_t(q)\gamma_t(q',q)\beta_{t+1}(q')}{\sum_{q=1}^{Q}\sum_{q'=1}^{Q}\alpha_t'(q)\gamma_t(q',q)\beta_{t+1}(q')}. \quad (83)$$

Define a diagonal matrix

$$D_t = \mathrm{diag}\{D_t(1),\cdots,D_t(Q)\}$$

where

$$D_t(q) = \begin{cases} \sum_{a\in\{-1,+1\}} \Pr(y_t|h_t=A_q,X_t=a)\Pr(X_t=a), \\ \qquad\qquad\qquad\qquad\qquad\qquad \text{if } x_t \text{ is unknown} \\ \Pr(y_t|h_t=A_q,x_t)\Pr(x_t), \quad\text{if } x_t \text{ is known.} \end{cases}$$

Define $\boldsymbol{\alpha}_t = [\alpha_t(1),\cdots,\alpha_t(Q)]^T$ and $\boldsymbol{\alpha}_t' = [\alpha_t'(1),\cdots,\alpha_t'(Q)]^T$. We can write the recursion formula (8) and (9) in matrix form as

$$\boldsymbol{\alpha}_t = PD_{t-1}PD_{t-2}\cdots PD_{t-m}\boldsymbol{\alpha}_{t-m}$$
$$\boldsymbol{\alpha}_t' = PD_{t-1}PD_{t-2}\cdots PD_{t-m}\boldsymbol{\alpha}_{t-m}'.$$

In the above, $\boldsymbol{\alpha}_{t-m} = \overline{\mathbf{p}}$ and $\boldsymbol{\alpha}_{t-m}' = PD_{t-m-1}\cdots PD_{t-m'}\overline{\mathbf{p}}$, where $P$ is the state-transition matrix and $\overline{\mathbf{p}} = [P_1,\cdots,P_Q]^T$ is the vector of stationary state probability. Since it is assumed that $P>0$, $D_t>0$, and $\boldsymbol{\pi}>0$, the forward state probability vectors are strictly positive $\boldsymbol{\alpha}_t > 0$ and $\boldsymbol{\alpha}_t' > 0$ for any $t$. Therefore, applying the following inequality

$$\frac{\sum_i u(i)}{\sum_i v(i)} \leq \max_i \frac{u(i)}{v(i)} \quad \text{where } u(i)>0,\, v(i)>0 \quad (84)$$

to (83) yields

$$\frac{\Lambda'(x_t)}{1+\Lambda'(x_t)}\frac{1+\Lambda(x_t)}{\Lambda(x_t)} \leq \max_{1\leq q\leq Q}\frac{\alpha_t'(q)}{\alpha_t(q)} \max_{1\leq q\leq Q}\frac{\alpha_t(q)}{\alpha_t'(q)}. \quad (85)$$

Take logarithm on both sides of (85) and apply the definition (26) of the Hilbert metric, we have

$$\ln\left(\frac{\Lambda'(x_t)}{1+\Lambda'(x_t)}\frac{1+\Lambda(x_t)}{\Lambda(x_t)}\right) \leq d(\boldsymbol{\alpha}_t,\boldsymbol{\alpha}_t'). \quad (86)$$

From the property of nonnegative matrices products [31] that the multiplication of a positive matrix $M>0$ and positive vectors $\mathbf{u}>0$ and $\mathbf{v}>0$ is a strict contraction with respect to the Hilbert metric, $d(M\mathbf{u},M\mathbf{v}) \leq \tau(M)d(\mathbf{u},\mathbf{v})$, the right hand side of (86) is upper bounded as

$$d(\boldsymbol{\alpha}_t,\boldsymbol{\alpha}_t') \leq \tau(PD_{t-1}PD_{t-2}\cdots PD_{t-m+1})$$
$$\times d(PD_{t-m}\boldsymbol{\alpha_{t-m}}, PD_{t-m}\boldsymbol{\alpha_{t-m}'})$$
$$\leq \tau(PD_{t-1})\tau(PD_{t-2})\cdots\tau(PD_{t-m+1})$$
$$\times d(PD_{t-m}\boldsymbol{\alpha_{t-m}}, PD_{t-m}\boldsymbol{\alpha_{t-m}'}) \quad (87)$$
$$= \tau(P)^{m-1}d(PD_{t-m}\boldsymbol{\alpha_{t-m}}, PD_{t-m}\boldsymbol{\alpha_{t-m}'}) \quad (88)$$

where (87) is due to the property of the Birkhoff contraction coefficient that $\tau(M_1M_2) \leq \tau(M_1)\tau(M_2)$ for $M_1>0$ and $M_2>0$, and (88) follows $\tau(M_1D) = \tau(M_1)$ for a diagonal matrix $D$ with positive diagonal entries.

In the following, we derive the upper bound for the term $d(PD_{t-m}\boldsymbol{\alpha_{t-m}}, PD_{t-m}\boldsymbol{\alpha_{t-m}'})$ in (88). Let $\mathbf{a} = [a(1),\cdots,a(Q)]^T$ and $\mathbf{a}' = [a'(1),\cdots,a'(Q)]^T$ be some positive vectors. By the definition (26), we have

$$d\Big(\sum_{i=1}^{Q} a(i)\mathbf{p}_i,\ a'(j)\mathbf{p}_j\Big)$$
$$= \ln\left(\max_q \frac{\sum_{i=1}^{Q} a(i)P(q,i)}{a'(j)P(q,j)} \left(\min_k \frac{\sum_{i=1}^{Q} a(i)P(k,i)}{a'(j)P(k,j)}\right)^{-1}\right)$$
$$\leq \ln\left(\sum_{i=1}^{Q} \max_q \frac{a(i)P(q,i)}{a'(j)P(q,j)} \left(\sum_{i=1}^{Q} \min_k \frac{a(i)P(k,i)}{a'(j)P(k,j)}\right)^{-1}\right) \quad (89)$$
$$\leq \ln\left(\max_i \left(\max_q \frac{P(q,i)}{P(q,j)} \left(\min_k \frac{P(k,i)}{P(k,j)}\right)^{-1}\right)\right) \quad (90)$$
$$= \max_i d(\mathbf{p}_i,\ \mathbf{p}_j) \quad (91)$$

where (89) is derived by moving the max and min operator inside the summation, (90) follows the inequality (84) and (91) follows the definition (26). Now, apply (91) twice, we have

$$d\Big(\sum_{i=1}^{Q} a(i)\mathbf{p}_i,\ \sum_{j=1}^{Q} a'(j)\mathbf{p}_j\Big) \leq \max_j d\Big(\sum_{i=1}^{Q} a(i)\mathbf{p}_i,\ \mathbf{p}_j\Big)$$
$$\leq \max_{i,j} d(\mathbf{p}_i,\ \mathbf{p}_j).$$

Hence

$$d(PD_{t-m}\boldsymbol{\alpha}_{t-m}, PD_{t-m}\boldsymbol{\alpha}'_{t-m}) \leq \max_{i,j} d(\mathbf{p}_i,\ \mathbf{p}_j). \quad (92)$$

Combining (82), (86), (88), and (92) completes the proof.

## APPENDIX II
## PROOF OF THE MONOTONICITY OF PILOT-UTILITY FUNCTION

For convenience, we re-write the definition of pilot-utility function (41) here

$$\mu(x) = \lim_{n\to\infty} I_{\mathbf{\Pi}}(X_t; \langle\mathbf{Y}\rangle_{t-n}^{t+n} | \langle\mathbf{Z}\rangle_{t-n}^{t-1}, \langle\mathbf{Z}\rangle_{t+1}^{t+n}) \quad (93)$$

where

$$Z_t = \begin{cases} X_t, & \text{with probability } x \\ \phi, & \text{with probability } 1-x \end{cases}$$

is the random training symbol. Let $0 \leq x \leq y \leq 1$. We define an additional sequence of training symbols $\overline{\mathbf{Z}}$, where

$$\overline{Z}_t = \begin{cases} \phi, & \text{if } Z_t = X_t \\ X_t, & \text{with probability } y-x \text{ if } Z_t = \phi \\ \phi, & \text{with probability } 1-(y-x) \text{ if } Z_t = \phi \end{cases}$$

and $\mathbf{Z}' = \mathbf{Z} \bigcup \overline{\mathbf{Z}}$. It can be shown that $Z'_t$ is i.i.d. with probability distribution $\Pr(Z'_t = X_t) = y$ and $\Pr(Z'_t = \phi) = 1-y$. Therefore

$$\mu(y) = \lim_{n\to\infty} I_{\mathbf{\Pi}}(X_t; \langle\mathbf{Y}\rangle_{t-n}^{t+n} | \langle\mathbf{Z}'\rangle_{t-n}^{t-1}, \langle\mathbf{Z}'\rangle_{t+1}^{t+n}). \quad (94)$$

From (93) and (94) and the chain rule and the non-negativity of the mutual information, we have

$$\mu(y) - \mu(x)$$
$$= \lim_{n\to\infty} I_{\mathbf{\Pi}}(X_t; \langle\overline{\mathbf{Z}}\rangle_{t-n}^{t-1}, \langle\overline{\mathbf{Z}}\rangle_{t+1}^{t+n} | \langle\mathbf{Y}\rangle_{t-n}^{t+n}, \langle\mathbf{Z}\rangle_{t-n}^{t-1}, \langle\mathbf{Z}\rangle_{t+1}^{t+n})$$
$$\geq 0. \quad (95)$$

**Teng Li** (S'05-M'07) received the B.S. degree from Shanghai Jiao Tong University, Shanghai, China, in 1999 and the M.S. and Ph.D. degrees in 2003, and 2007, respectively, from the University of Notre Dame, Notre Dame, IN, all in electrical engineering.

Since 2008, he has been with Augusta Technology USA, Inc, where he works on the research and development in the field of wireless communications. He was a senior design engineer in the signal processing division at Marvell Semiconductor, Inc., Santa Clara, CA from 2006 to 2008. From 1999 to 2000, he was a software engineer in Ericsson Communication Software Research and Development Company, China. His research interests include information theory, coding, signal processing for communications, and distributed sensor networks.

**Oliver M. Collins** (S'88-SM'99-F'02) was born in Washington, DC. He received the B.S. degree in engineering and applied science, and the M.S. and Ph.D. degrees in electrical engineering in 1986, 1987, and 1989, respectively, all from the California Institute of Technology, Pasadena. From 1989 to 1995, he was an Assistant Professor and then an Associate Professor with the Department of Electrical and Computer Engineering, Johns Hopkins University, Baltimore, MD. In September 1995, he became an Associate Professor with the Department of Electrical Engineering, University of Notre Dame, Notre Dame, IN. He became a Full Professor in 2001 and currently teaches courses in communications, information theory, coding, and complexity theory. Dr. Collins was the recipient of the 1994 IEEE Thompson Prize Paper Award, the 1994 Marconi Young Scientist Award presented by the Marconi Foundation, and the 1998 IEEE Judith Resnik Award.

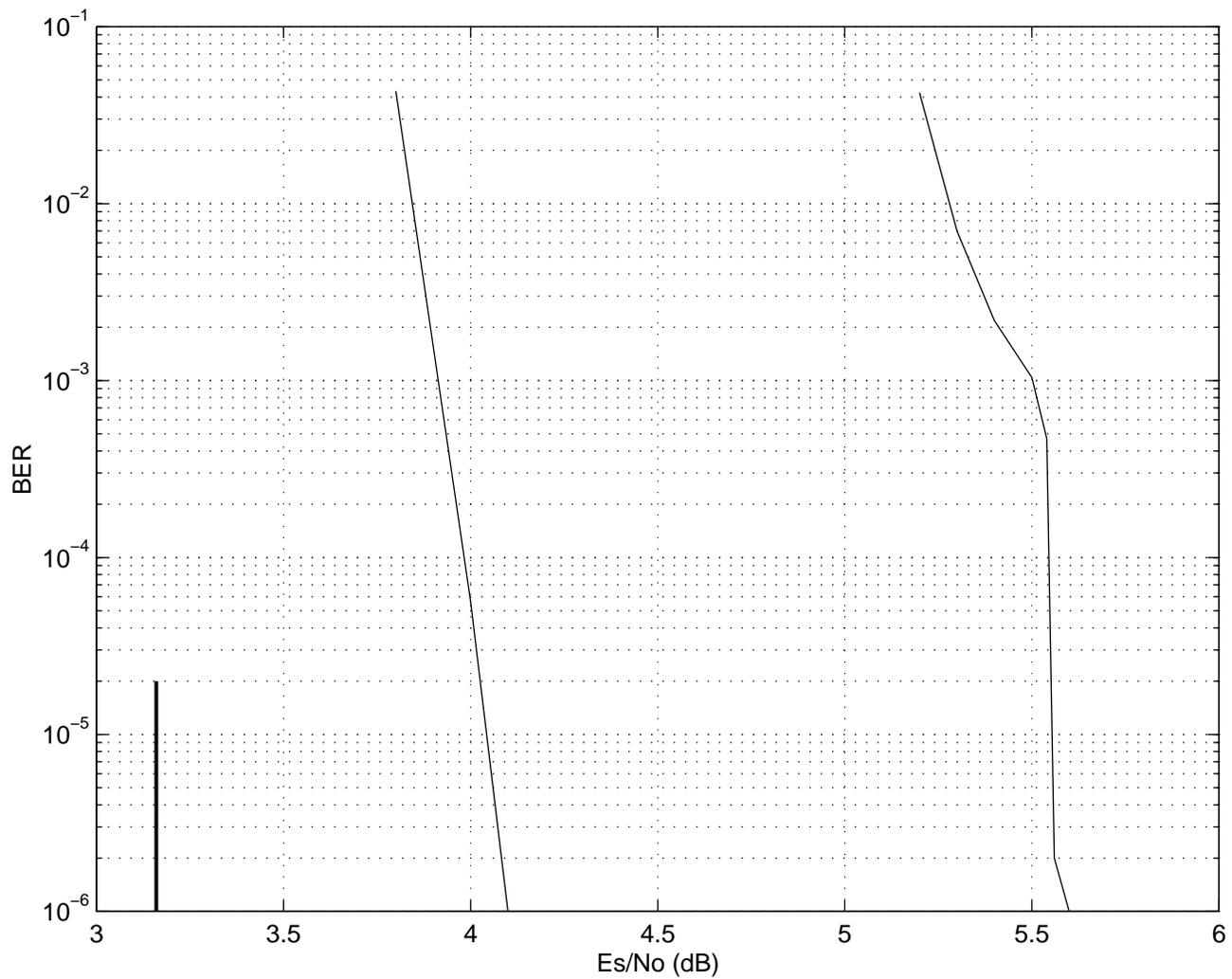

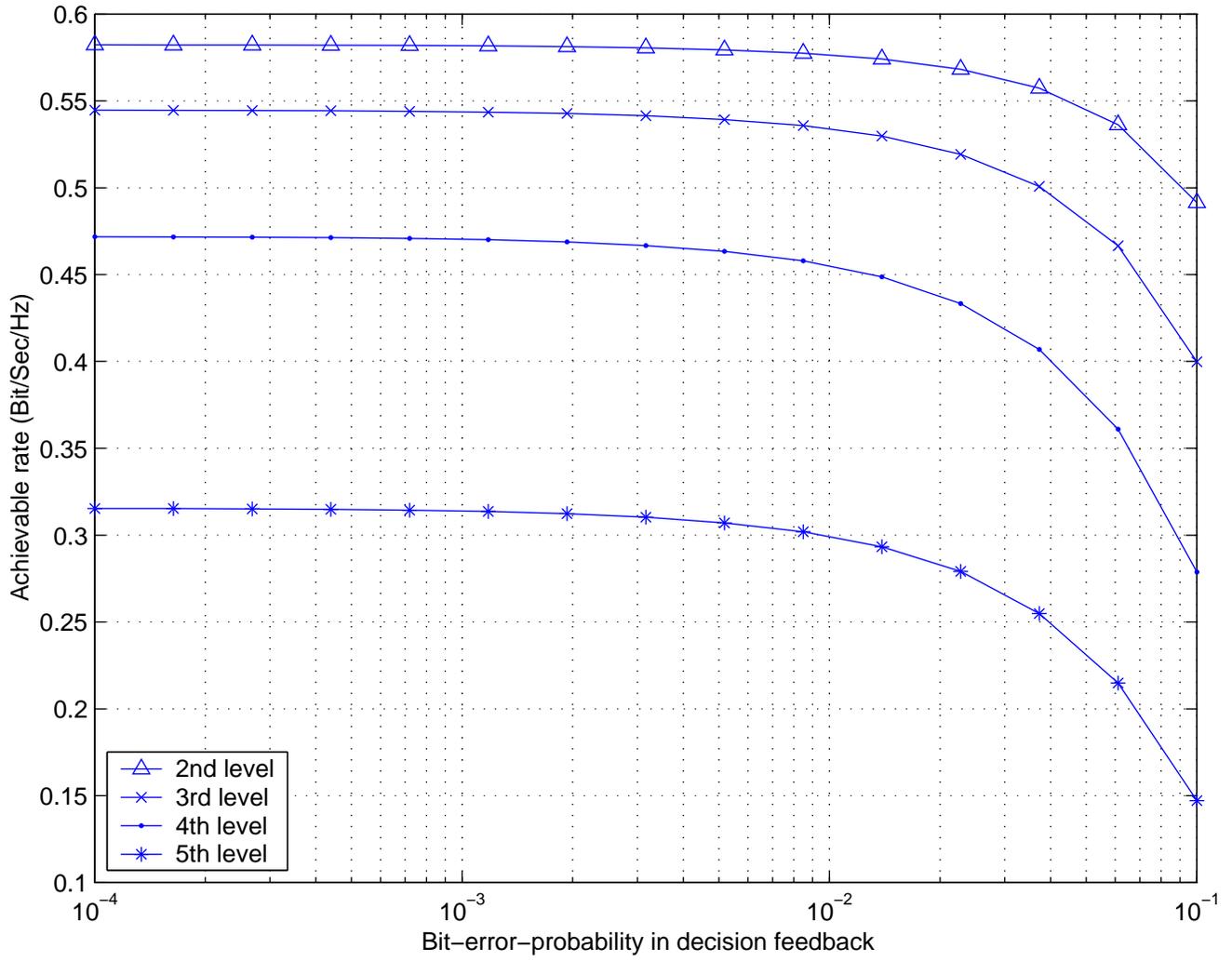

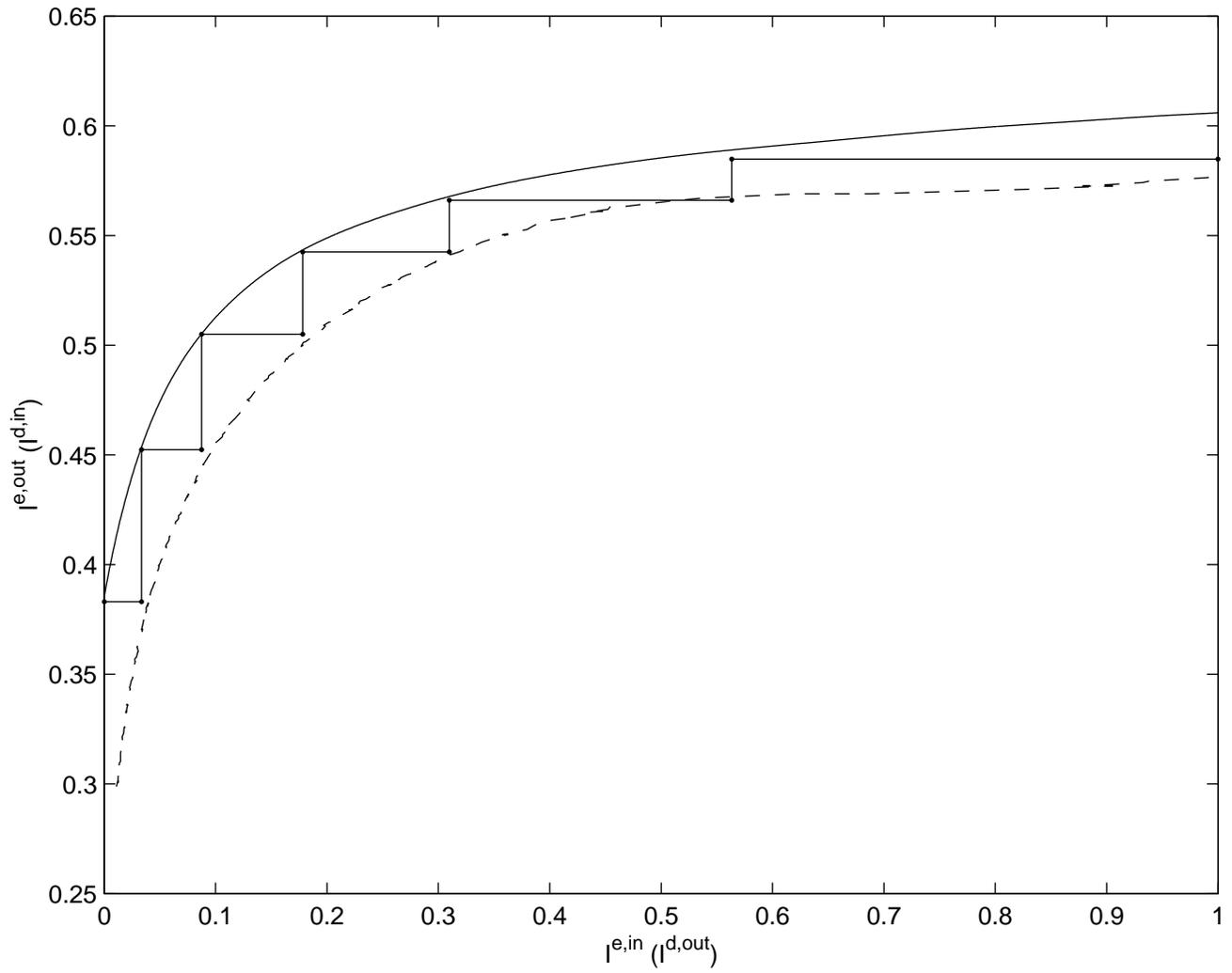